\documentclass[AMS,STIX2COL]{WileyNJD-v2}
\usepackage[]{graphicx}
\usepackage{amsmath}
\usepackage{amssymb}
\usepackage{caption}
\usepackage{mathtools}
\usepackage{changes}
\usepackage{multirow}
\usepackage{verbatim}
\usepackage{mathrsfs}
\usepackage{hyperref}
\articletype{Regular Paper}%

\received{26 April 2016}
\revised{6 June 2016}
\accepted{6 June 2016}

\begin{document}

\title{Adaptive Frequency-limited $\mathcal{H}_2$-Model Order Reduction}

\author[1]{Umair Zulfiqar}

\author[2]{Victor Sreeram}

\author[1,3,4]{Xin Du*}

\authormark{UMAIR ZULFIQAR \textsc{et al}}

\address[1]{\orgdiv{School of Mechatronic Engineering and Automation, and Shanghai Key Laboratory of Power Station Automation Technology}, \orgname{Shanghai University}, \orgaddress{\state{Shanghai}, \country{China}}}
\address[2]{\orgdiv{Department of Electrical, Electronics and Computer Engineering}, \orgname{The University of Western Australia}, \orgaddress{\state{WA}, \country{Australia}}}
\address[3]{\orgdiv{Key Laboratory of Knowledge Automation for Industrial Processes}, \orgname{Ministry of Education}, \orgaddress{\state{Beijing}, \country{China}}}

\address[4]{\orgdiv{Key Laboratory of Modern Power System Simulation and Control \& Renewable Energy Technology}, \orgname{Ministry of Education (Northeast Electric Power University)}, \orgaddress{\state{Jilin}, \country{China}}}

\corres{*Xin Du, School of Mechatronic Engineering and Automation, and Shanghai Key Laboratory of Power Station Automation Technology, Shanghai University. \email{umair.zulfiqar@research.uwa.edu.au}}

\presentaddress{No. 99 Shangda Road, Baoshan District, Shanghai 200444 China.}

\abstract[Summary]{In this paper, we present an adaptive framework for constructing a pseudo-optimal reduced model for the frequency-limited $\mathcal{H}_2$-optimal model order reduction problem. We show that the frequency-limited pseudo-optimal reduced-order model has an inherent property of monotonic decay in error if the interpolation points and tangential directions are selected appropriately. We also show that this property can be used to make an automatic selection of the order of the reduced model for an allowable tolerance in error. The proposed algorithm adaptively increases the order of reduced model such that the frequency-limited $\mathcal{H}_2$-norm error decays monotonically irrespective of the choice of interpolation points and tangential directions. The stability of the reduced-order model is also guaranteed. Additionally, it also generates the approximations of the frequency-limited system Gramians that monotonically approach the original solution. Further, we show that the low-rank alternating direction implicit iteration method for solving large-scale frequency-limited Lyapunov equations implicitly performs frequency-limited pseudo-optimal model order reduction. We consider two numerical examples to validate the theory presented in the paper.}

\keywords{$\mathcal{H}_2$-norm, cumulative framework, limited frequency interval, pseudo-optimal}
\maketitle
\section{Introduction}\label{sec1}
Model order reduction (MOR) is a process of obtaining a reduced-order model (ROM) that accurately preserves the input-output dynamics and some desired properties of the original high-order model. The specific properties and dynamics to be preserved lead to various classes of the MOR procedures. In projection-based MOR techniques, the original high-order model is projected onto a reduced subspace to obtain a ROM that preserves the desired and dominant characteristics of the original model. The ROM can then be used as a surrogate in the design and analysis to provide significant numerical and computational advantages \cite{benner2005dimension,benner2017model,chow2013power,obinata2012model,schilders2008model}.\\
Moore presented one of the most widely used MOR techniques in \cite{moore1981principal}. Moore's balanced truncation (BT) truncates the states that have an insignificant share in the overall energy transfer, and thus it provides a compact ROM that contains the important states (and dynamics) of the original model. In many applications, a specific frequency interval is more important, i.e., the ROM should maintain a superior accuracy within that desired frequency interval. Gawronski and Juang generalized BT for the frequency-limited MOR scenario in \cite{gawronski1990model}. In frequency-limited BT (FLBT), the states that have an insignificant share in the energy transfer within the desired frequency interval are truncated. BT and FLBT require the solutions of large-scale dense Lyapunov equations that become computationally infeasible when the original model is a large-scale model. In \cite{balakrishnan2001efficient,benner2016frequency,kurschner2018balanced}, their applicability is extended to large-scale systems by using low-rank approximate solutions of the Lyapunov equations that can be efficiently computed. The stability of the ROM is not guaranteed in FLBT. The so-called modified FLBT algorithms presented in the literature like \cite{ghafoor2008survey,imran2015frequency,imran2017frequency,toor2018improved} fix this problem and guarantee the stability of the ROM. However, these modifications generally offer poor approximation and also make FLBT an even more computationally expensive algorithm. Reference \cite{zulfiqar2016new} heuristically suggests a modification wherein the large-scale Lyapunov equations are solved several times until a stable ROM is obtained by adjusting a user-defined perturbation. Such an approach is only viable for small-scale problems due to the excessive computational cost associated with each iteration. The applicability of FLBT to large-scale systems is only viable if the Lyapunov equations are solved approximately using the low-rank approximation methods like \cite{benner2016frequency}. Several other generalizations have also been proposed in the literature like \cite{haider2018frequency,imran2015model,imran2016model,jazlan2016frequency,shaker2013generalized,shaker2006frequency,zulfiqar2020time} to extend the applicability of FLBT to more general classes of linear and bilinear systems.

Moment matching is another important class of MOR techniques. In moment matching, a ROM is constructed that interpolates the original transfer function at some selected frequency points. This can be achieved by using Krylov subspace-based algorithms that are computationally efficient. Unlike BT, the moment matching algorithms do not require any solution of large-scale Lyapunov equations and thus can handle large-scale systems \cite{beattie2014model}. In \cite{gugercin2008h_2,van2008h2}, the $\mathcal{H}_2$-optimal MOR problem \cite{wilson1970optimum} is expressed as a tangential interpolation problem. This allows the usage of a rational Krylov subspace-based framework for finding a local optimum for this problem. In \cite{gugercin2008h_2}, an iterative rational Krylov algorithm (IRKA) is proposed that constructs a local optimum for the $\mathcal{H}_2$-optimal MOR problem upon convergence. The original algorithm was presented for single-input single-output (SISO) systems and was later generalized for multi-input multi-output (MIMO) systems in \cite{van2008h2}. The accuracy of the ROM constructed by IRKA is comparable to that of BT, and thus it is considered as a gold standard among moment matching techniques. However, unlike BT, the stability of the ROM is not guaranteed. IRKA is an iterative framework that starts with a random selection of the interpolation points and tangential directions. It generally converges quickly for the SISO case, even if the interpolation points and tangential directions are chosen arbitrarily. The speed of convergence, however, slows down as the number of inputs and outputs is increased. There are some trust-region methods reported in the literature that speed up the convergence of IRKA like \cite{beattie2009trust}.

In \cite{wolf2014h}, an iteration-free algorithm is proposed that satisfies a subset of the first-order optimality conditions that IRKA satisfies upon convergence. The algorithm is named as ``Pseudo-optimal Rational Krylov (PORK)" algorithm. The stability of the ROM is also guaranteed in PORK. In \cite{panzer2014model}, a cumulative reduction (CURE) scheme for moment matching is proposed that generates the ROM in steps, and the final ROM is the accumulation of all the interim ROMs generated at each step. The interpolation conditions induced at each step are retained and accumulated in the final ROM. CURE has an interesting property that if PORK is used to generate the ROM at each step, the $\mathcal{H}_2$-norm error continues to decay monotonically irrespective of the choice of interpolation points and the tangential directions. Moreover, the final ROM is also a pseudo-optimal ROM, i.e., it satisfies a subset of the first-order optimality conditions \cite{wilson1970optimum} for the $\mathcal{H}_2$-optimal MOR problem. The monotonic decay in error is an important property as it can make an adaptive choice of the order of the ROM possible. Though the error generally decreases when the order is increased in most of the MOR techniques, it often increases as well. This hinders an adaptive choice of the order as there is no assurance that the error decays by constructing a higher-order ROM. Moreover, the ROMs constructed in the previous steps of CURE are accumulated and reused, which saves a great deal of the computational cost.

In \cite{petersson2014model}, the first-order optimality conditions for the frequency-limited $\mathcal{H}_2$-optimal MOR are derived, and a nonlinear optimization algorithm is presented to achieve these conditions. The algorithm presented in \cite{petersson2014model} is not feasible in a large-scale setting due to its high computational cost. The optimality conditions for the frequency-limited $\mathcal{H}_2$-MOR are expressed as Hermite interpolation conditions in \cite{vuillemin2014frequency}, and a descent-based optimization algorithm is presented to achieve these conditions. This algorithm is also not feasible in a large-scale setting. In \cite{vuillemin2013h2}, IRKA is heuristically generalized to the frequency-limited scenario. Frequency-limited IRKA (FLIRKA) \cite{vuillemin2013h2} generates a high-fidelity ROM, even when it does not converge. It is a computationally efficient algorithm; however, it does not satisfy the first-order optimality conditions for the frequency-limited $\mathcal{H}_2$-optimal-MOR problem. In \cite{zulfiqar2020frequency}, an iteration-free and computationally efficient algorithm is presented that constructs a ROM, which satisfies a subset of the first-order optimality conditions for the frequency-limited $\mathcal{H}_2$-optimal-MOR problem. This algorithm is a generalization of PORK, and it is named frequency-limited PORK (FLPORK). The theoretical connection between FLIRKA and FLPORK is also investigated in \cite{zulfiqar2020frequency}. The stability of the ROM is guaranteed in FLPORK.

In this paper, we first show a few properties of FLPORK that were not recognized in \cite{zulfiqar2020frequency}. In particular, we show that the Krylov subspace of (the infinite frequency interval) moment matching methods is related to the Krylov subspace of frequency-limited moment matching. We further show that FLPORK maintains a particular structure in the ROM, and FLIRKA \cite{vuillemin2013h2} fails to satisfy any optimality condition because it cannot preserve this structure. We also explain why FLIRKA still manages to ensure high-fidelity. We further show that FLPORK has an inherent property of monotonic decay in error if the selection of interpolation data satisfies a particular condition. We modify CURE \cite{panzer2014model} such that if the ROM accumulated at each step is obtained using FLPORK, the frequency-limited $\mathcal{H}_{2}$-norm error continues to decay monotonically irrespective of the choice of interpolation points and the tangential directions. Additionally, the proposed algorithms also provide an approximation of frequency-limited Gramians in a computationally efficient way that can be used to extend the applicability of FLBT to large-scale systems. The approximate Gramians monotonically approach the original solution irrespective of the choice of interpolation points and the tangential directions. We also show that these Gramians are related to the alternating implicit iteration (ADI) method, and the low-rank ADI (LR-ADI) method implicitly performs frequency-limited pseudo-optimal MOR. We also propose an adaptive scheme for the selection of interpolation data, as well as the order of the ROM. The proposed algorithm automatically constructs the ROM without any user interference once the desired frequency interval and allowable tolerance in error are supplied. We establish the significance of the proposed algorithm by demonstrating its effectiveness on two numerical examples.
\section{Preliminaries}
Consider a $n^{th}$ order linear time-invariant system with the following state-space realization
\begin{align}
\dot{x}(t)=Ax(t)+Bu(t),&&y(t)=Cx(t)+Du(t),\label{eq:1}
\end{align} where $x(t)\in\mathbb{R}^{n\times 1}$, $u(t)\in\mathbb{R}^{m\times 1}$, and $y(t)\in\mathbb{R}^{p\times 1}$ are state, input, and output vectors, respectively. $A\in\mathbb{R}^{n\times n}$, $B\in\mathbb{R}^{n\times m}$, $C\in\mathbb{R}^{p\times n}$, and $D\in\mathbb{R}^{p\times m}$. The system (\ref{eq:1}) has the following transfer function representation
\begin{align}
H(s)=C(sI-A)^{-1}B+D.\nonumber
\end{align}
The MOR problem is to construct a $r^{th}$-order model that approximates the original high-order model (\ref{eq:1}) where $r\ll n$. Let the ROM have the following state-space and transfer function representations
\begin{align}
\dot{x}_r(t)&=A_rx_r(t)+B_ru(t),\hspace{1cm}y_r(t)=C_rx_r(t)+D_ru(t),\nonumber\\
H_r(s)&=C_r(sI-A_r)^{-1}B_r+D_r,\nonumber
\end{align} where $A_r\in\mathbb{R}^{r\times r}$, $B_r\in\mathbb{R}^{r\times m}$, $C_r\in\mathbb{R}^{p\times r}$, and $D_r\in\mathbb{R}^{p\times m}$. MOR aims to ensure that $||H(s)-H_r(s)||$ is small in some defined sense depending on the nature of the problem. In the frequency-limited MOR problem, a ROM is sought that ensures a small frequency domain error $||H(j\omega)-H_r(j\omega)||$ within the desired frequency interval $\Omega=[\omega_1,\omega_2]$.

Let $V_r$ and $W_r$ be the input and output reduction matrices, respectively, of the projection-based MOR. Then these project the original high-order model onto a reduced subspace as the following
\begin{align}
A_r=W_r^TAV_r,\hspace*{2mm} B_r=W_r^TB,\hspace*{2mm} C_r=CV_r, \hspace*{2mm} D_r=D,\nonumber
\end{align} where $V_r\in \mathbb{R}^{n\times r}$ and $W_r\in \mathbb{R}^{n\times r}$.

Let $P_\Omega$ be the frequency-limited controllability Gramian and $Q_\Omega$ be the frequency-limited observability Gramian of the state-space realization (\ref{eq:1}) within the desired frequency interval $[\omega_1,\omega_2]$. $P_\Omega$ and $Q_\Omega$ solve the following Lyapunov equations
\begin{align}
AP_\Omega+P_\Omega A^T+\mathscr{F}(A)BB^T+BB^T\mathscr{F}(A^T)=0,\label{e2}\\
A^TQ_\Omega+Q_\Omega A+\mathscr{F}(A^T)C^TC+C^TC\mathscr{F}(A)=0,\label{e3}
\end{align} where
\begin{align}
\mathscr{F}(A)&=\frac{1}{2\pi}\int_{-\omega_2}^{\omega_2}(j\nu I-A)^{-1}d\nu-\frac{1}{2\pi}\int_{-\omega_1}^{\omega_1}(j\nu I-A)^{-1}d\nu\nonumber\\
&=\frac{1}{2\pi}ln\big((j\omega_2I+A)(-j\omega_2I+A)^{-1}\big)-\nonumber\\
&\hspace*{2cm}\frac{1}{2\pi}ln\big((j\omega_1I+A)(-j\omega_1I+A)^{-1}\big).\label{e4}
\end{align}
 The frequency-limited $\mathcal{H}_2$-norm \cite{petersson2013nonlinear,petersson2014model} ($\mathcal{H}_{2,\Omega}$-norm) of $H(s)$ is given by
\begin{align}
||H&(s)||_{\mathcal{H}_{2,\Omega}}\nonumber\\
&=\Big[tr\Big(CP_\Omega C^T+2\big(C\mathscr{F}(A)B+D\frac{\omega_2-\omega_1}{2\pi}\big)D^T\Big)\Big]^{\frac{1}{2}}\nonumber\\
&=\Big[tr\Big(B^TQ_\Omega B+2\big(C\mathscr{F}(A)B+D\frac{\omega_2-\omega_1}{2\pi}\big)D^T\Big]^{\frac{1}{2}}.\nonumber
\end{align}
When the frequency interval is specified as $[-\infty,\infty]$, $P_\Omega=P$, $Q_\Omega=Q$, and $||\cdot||_{\mathcal{H}_{2,\Omega}}=||\cdot||_{\mathcal{H}_2}$ where $P$, $Q$, and $\mathcal{H}_2$ are the standard controllability Gramian, observability Gramian, and $\mathcal{H}_2$-norm, respectively. In the frequency-limited $\mathcal{H}_2$-MOR, a ROM is sought that minimizes $||H(s)-H_r(s)||^2_{\mathcal{H}_{2,\Omega}}$. Table \ref{tab0} enlists the important mathematical notations used in the text.
\subsection{Necessary Conditions for Local Optimum}
Let $(A_e,B_e,C_e,D_e)$ be a state-space realization of $H(s)-H_r(s)$ wherein
\begin{align}
A_e&=\begin{bmatrix}A&0\\0&A_r\end{bmatrix}, &B_e&=\begin{bmatrix}B\\B_r\end{bmatrix},\nonumber\\
C_e&=\begin{bmatrix}C&-C_r\end{bmatrix}, &D_e&=D-D_r.\label{r2.e5}
\end{align}
Let us denote the frequency-limited controllability Gramian and the frequency-limited observability Gramian of $(A_e,B_e,C_e,D_e)$ as $P_{e,\Omega}$ and $Q_{e,\Omega}$, respectively. Then $P_{e,\Omega}$ and $Q_{e,\Omega}$ can be partitioned according to (\ref{r2.e5}) as
\begin{align}
P_{e,\Omega}&=\begin{bmatrix}P_\Omega&\hat{P}_\Omega\\\hat{P}_\Omega^T&P_{r,\Omega}\end{bmatrix}&\textnormal{and}&&Q_{e,\Omega}&=\begin{bmatrix}Q_\Omega&\hat{Q}_\Omega\\\hat{Q}_\Omega^T&Q_{r,\Omega}\end{bmatrix}\nonumber
\end{align} where $\hat{P}_\Omega$, $P_{r,\Omega}$, $\hat{Q}_\Omega$, and $Q_{r,\Omega}$ satisfy the following Lyapunov equations
\begin{align}
A\hat{P}_\Omega+\hat{P}_\Omega A_r^T+\mathscr{F}(A)BB_r^T+BB_r^T\mathscr{F}(A_r^T)=0,\label{r2.eq.6}\\
A_rP_{r,\Omega}+P_{r,\Omega} A_r^T+\mathscr{F}(A_r)B_rB_r^T+B_rB_r^T\mathscr{F}(A_r^T)=0,\label{r2.eq.7}
\end{align}
\begin{align}
A^T\hat{Q}_\Omega+\hat{Q}_\Omega A_r-\mathscr{F}(A^T)C^TC_r-C^TC_r\mathscr{F}(A_r)=0,\label{r2.eq.8}\\
A_r^TQ_{r,\Omega}+Q_{r,\Omega} A_r+\mathscr{F}(A_r^T)C_r^TC_r+C_r^TC_r\mathscr{F}(A_r)=0.\label{r2.eq.9}
\end{align}
Further, let $P$ and $\hat{P}$ satisfy the following Lyapunov equations
\begin{align}
AP+PA^T+BB^T&=0,\nonumber\\
A\hat{P}+\hat{P}A_r^T+BB_r^T&=0.\nonumber
\end{align}
Then the local optimum of $||H(s)-H_r(s)||^2_{\mathcal{H}_{2,\Omega}}$ satisfies the following Gramians-based conditions \cite{petersson2013nonlinear,petersson2014model}
\begin{align}
\hat{Q}_{\Omega}^T\hat{P}+Q_{r,\Omega}P_r-R_1=0,\\
\hat{Q}_{\Omega}^TB+Q_{r,\Omega}B_r-\mathscr{F}(A_r)C_r^T\big(D-D_r\big)=0,\label{r2.e7}\\
C\hat{P}_\Omega-C_rP_{r,\Omega}+\big(D-D_r\big)B_r^T\mathscr{F}(A_r^T)=0,\label{r2.e8}\\
C\mathscr{F}(A)B+\big(D-D_r\big)\frac{\omega_2-\omega_1}{\pi}-C_r\mathscr{F}(A_r)B_r,
\end{align} where
\begin{align}
R_1&=Re\big(\mathcal{L}(-A_r-j\omega_2I,R_2)-\mathcal{L}(-A_r-j\omega_1I,R_2)\big)^T,\nonumber\\
R_2&=C_r^TC_rP_r-C_r^TC\hat{P}-C_r^T(D-D_r)B_r^T.\nonumber
\end{align}
When either (\ref{r2.e7}) or (\ref{r2.e8}) is satisfied, the following holds
\begin{align}
||H(s)-H_r(s)||_{\mathcal{H}_{2,\Omega}}^2=||H(s)||_{\mathcal{H}_{2,\Omega}}^2-||H_r(s)||_{\mathcal{H}_{2,\Omega}}^2.\label{r2.e10}
\end{align}
\begin{table}[!t]
\centering
\caption{Mathematical Notations}\label{tab0}
\begin{tabular}{|c|c|}
\hline
Notation & Meaning \\ \hline
$Re(\cdot)$   & Real part of the matrix\\
$\bar{[\cdot]}$   & Conjugate of the matrix\\
$tr(\cdot)$  & Trace of the matrix\\
$[\cdot]^*$& Hermitian of the matrix\\
$\mathcal{L}(\cdot)$& Fr\'{e}chet derivative of matrix logarithm\\
$\oplus$& Direct sum of the matrices\\
$Ran(\cdot)$ & Range of the matrix\\
$\underset {i=1,\cdots,r}{span}\{\cdot\}$ & Span of the set of $r$ vectors\\\hline
\end{tabular}
\end{table}
We refer to the ROM that satisfies (\ref{r2.e10}) as the frequency-limited pseudo-optimal ROM. The optimal choice of $D_r$ is determined by the optimal choices of $A_r$, $B_r$, and $C_r$ and is not involved in the optimization process. Therefore, we assume that $D=0$ and $D_r=0$ for the remainder of this paper without loss of any generality. In case $D$ is nonzero, $D_r$ can be set as $D_r=D$ without changing any result presented in the remainder of this paper. Let $H(s)$ and $H_r(s)$ have simple poles and have the following pole-residue forms
\begin{align}
H(s)=\sum_{i=1}^{n}\frac{l_ir_i^T}{s-\lambda_i}\textnormal{ and }H_r(s)=\sum_{i=1}^{r}\frac{\tilde{l}_i\tilde{r}_i^T}{s-\tilde{\lambda}_i}.\nonumber
\end{align} Define $T_\Omega(s)$ and $T_{r,\Omega}(s)$ as
\begin{align}
T_\Omega(s)&=H_\Omega(s)+H(s)\mathscr{F}(-s),\nonumber\\
T_{r,\Omega}(s)&=H_{r,\Omega}(s)+H_r(s)\mathscr{F}(-s),\nonumber
\end{align}where
\begin{align}
H_\Omega(s)&=\sum_{i=1}^{n}\frac{l_ir_i^T}{s-\lambda_i}\mathscr{F}(\lambda_i),&&H_{r,\Omega}(s)&=\sum_{i=1}^{r}\frac{\tilde{l}_i\tilde{r}_i^T}{s-\tilde{\lambda}_i}\mathscr{F}(\tilde{\lambda}_i).\nonumber
\end{align}
Then the local optimum of $||H(s)-H_r(s)||^2_{\mathcal{H}_{2,\Omega}}$ satisfies the following bi-tangential Hermite interpolation conditions
\begin{align}
\tilde{l}_i^TT_\Omega^\prime(-\tilde{\lambda}_i)\tilde{r}_i&=\tilde{l}_i^TT_{r,\Omega}^\prime(-\tilde{\lambda}_i)\tilde{r}_i,\\
\tilde{l}_i^TT_\Omega(-\tilde{\lambda}_i)&=\tilde{l}_i^TT_{r,\Omega}(-\tilde{\lambda}_i),\label{r2.e12}\\
T_\Omega(-\tilde{\lambda}_i)\tilde{r}_i&=T_{r,\Omega}(-\tilde{\lambda}_i)\tilde{r}_i.\label{r2.e13}
\end{align}
\subsection{Literature Review}
In this subsection, some important MOR algorithms in the existing literature, which are most relevant to the results presented in this paper, are briefly surveyed.
\subsubsection{PORK \cite{wolf2014h}}
It is shown in \cite{gugercin2008h_2,van2008h2} that the local optimum of $||H(s)-H_r(s)||^2_{\mathcal{H}_2}$ satisfies the following bi-tangential Hermite interpolation conditions
\begin{align}
H(-\tilde{\lambda}_i)\tilde{r}_i&=H_r(-\tilde{\lambda}_i)\tilde{r}_i,\label{2}\\
\tilde{l}_i^TH(-\tilde{\lambda}_i)&=\tilde{l}_i^TH_r(-\tilde{\lambda}_i),\label{3}\\
\tilde{l}_i^TH^{\prime}(-\tilde{\lambda}_i)\tilde{r}_i&=\tilde{l}_i^TH_r^{\prime}(-\tilde{\lambda}_i)\tilde{r}_i.
\end{align}
When $H_r(s)$ satisfies either (\ref{2}) or (\ref{3}), the following holds
\begin{align}
||H(s)-H_r(s)||^2_{\mathcal{H}_2}=||H(s)||^2_{\mathcal{H}_2}-||H_r(s)||^2_{\mathcal{H}_2}.\label{5}
\end{align}
We refer to $H_r(s)$ as a pseudo-optimal ROM if it satisfies (\ref{5}). A pseudo-optimal ROM that satisfies (\ref{2}) can be generated by PORK as the following. Let $V_r$ satisfy the following condition
\begin{align}
Ran(V_r)=\underset {i=1,\cdots,r}{span}\{(\sigma_iI-A)^{-1}Bb_i\},\label{e9}
\end{align} where $\{\sigma_1,\cdots,\sigma_r\}$ and $\{b_1,\cdots,b_r\}$ are the interpolation points and the associated right tangential directions, respectively. Then $V_r$ enforces the interpolation condition $H(\sigma_i)b_i=H_r(\sigma_i)b_i$ for any $W_r$ such that $W_r^TV_r=I$. Define the oblique projection $\Pi=V_rW_r^T$ wherein $W_r$ is arbitrary. Further, define $B_\bot$, $L_r$, and $S_r$ as
\begin{align}
B_\bot&=(I-\Pi)B, &&L_r=(B_\bot^TB_\bot)^{-1}B_\bot^T\big(I-\Pi)AV_r,\nonumber\\
S_r&=W_r^T(AV_r-BL_r).\nonumber
\end{align}
Then $V_r$ solves the following Sylvester equation
\begin{align}
AV_r-V_rS_r-BL_r=0,\label{6}
\end{align} where the interpolation points are the eigenvalues of $S_r$ and the right tangential directions are encoded in $L_r$. Further, $A_r$ can be parameterized in $B_r$ without affecting the interpolation condition induced by $V_r$ as $A_r=S_r+B_rL_r$. Let $Q_s$ be the observability Gramian of the pair $(-S_r,L_r)$ that solves the following Lyapunov equation
\begin{align}
-S_r^TQ_s-Q_sS_r+L_r^TL_r=0.\nonumber
\end{align}
If $B_r$ is selected as $B_r=-Q_s^{-1}L_r^T$, the ROM satisfies (\ref{2}), i.e.,
\begin{align}
A_r&=-Q_s^{-1}S_r^TQ_s,& B_r&=-Q_s^{-1}L_r^T,&C_r&=CV_r,\nonumber
\end{align} Moreover, $P_r=Q_s^{-1}$ is the controllability Gramian of the pair $(A_r,B_r)$. Further, note that $V_r$ also satisfies the following Sylvester equation
\begin{align}
AV_r-V_rA_r-B_\bot L_r=0.\label{speq}
\end{align}This can readily be verified by multiplying the equation (\ref{speq}) by $W_r$ from the left. Similarly, a $W$-type PORK also exists, which ensures that $H_r(s)$ satisfies (\ref{3}); see \cite{wolf2014h} for details.
\subsubsection{CURE \cite{panzer2014model}}
CURE is an adaptive framework wherein $H_r(s)$ is constructed adaptively in $k$ steps such that at each step, new interpolation conditions are added without disturbing the previous ones. Let $\tilde{V}^{(i)}$, $\tilde{S}^{(i)}$, $\tilde{L}^{(i)}$, $(\tilde{A}^{(i)},\tilde{B}^{(i)},\tilde{C}^{(i)})$, and $B_{\perp}^{(i)}$ be the matrices of $i^{th}$ step for any $\tilde{W}^{(i)}$ such that $(\tilde{W}^{(i)})^T\tilde{V}^{(i)}=I$, i.e.,
\begin{align}
\tilde{A}^{(i)}&=(\tilde{W}^{(i)})^TA\tilde{V}^{(i)},&& \tilde{B}^{(i)}=(\tilde{W}^{(i)})^TB_{\perp}^{(i-1)},\nonumber\\
\tilde{C}^{(i)}&=C\tilde{V}^{(i)}\nonumber
\end{align} where
\begin{align}
A\tilde{V}^{(i)}-\tilde{V}^{(i)}\tilde{S}^{(i)}-B_{\perp}^{(i-1)}\tilde{L}^{(i)}=0,\nonumber
\end{align} $B_{\perp}^{(0)}=B$, and $B_{\perp}^{(i)}=B_{\perp}^{(i-1)}-\tilde{V}^{(i)}\tilde{B}^{(i)}$.\\
In CURE, the ROM is accumulated, and the order of the ROM keeps on growing after each iteration. The cumulative ROM and its associated matrices for $i=1,\cdots,k$ are computed as
\begin{align}
A_{c}^{(i)}&=\begin{bmatrix}A_{c}^{(i-1)}&0\\&\\\tilde{B}^{(i)}L_{c}^{(i-1)}&\tilde{A}^{(i)}\end{bmatrix},\hspace*{0.65cm} B_{c}^{(i)}=\begin{bmatrix}B_{c}^{(i-1)}\\\\\tilde{B}^{(i)}\end{bmatrix},\nonumber\\
C_{c}^{(i)}&=\begin{bmatrix}C_{c}^{(i-1)}&&\tilde{C}^{(i)}\end{bmatrix},\hspace*{1cm}L_{c}^{(i)}=\begin{bmatrix}L_{c}^{(i-1)}&&\tilde{L}^{(i)}\end{bmatrix},\nonumber\\
S_{c}^{(i)}&=\begin{bmatrix}S_{c}^{(i-1)}&-B_{c}^{(i-1)}\tilde{L}^{(i)}\\&\\0&\tilde{S}^{(i)}\end{bmatrix},\hspace*{0.1cm}V_{c}^{(i)}=\begin{bmatrix}V_{c}^{(i-1)}& \tilde{V}^{(i)}\end{bmatrix}\nonumber
\end{align} where
\begin{align}
AV_{c}^{(i)}-V_{c}^{(i)}S_{c}^{(i)}-BL_{c}^{(i)}=0,\nonumber
\end{align} and $A_{c}^{(0)}$, $B_{c}^{(0)}$, $C_{c}^{(0)}$, $L_{c}^{(0)}$, $S_{c}^{(0)}$, and $V_{c}^{(0)}$ are all empty matrices. Let $H_{c}^{(i)}\big(s\big)$ be the transfer function of the cumulative ROM, i.e.,
\begin{align}
H_{c}^{(i)}\big(s\big)=C_{c}^{(i)}\big(sI-A_{c}^{(i)}\big)^{-1}B_{c}^{(i)}.\nonumber
\end{align}

If $(\tilde{A}^{(i)},\tilde{B}^{(i)},\tilde{C}^{(i)})$ is computed by using PORK for $i=1,\cdots,k$, $H_{c}^{(i)}\big(s\big)$ stays pseudo-optimal, and $||H(s)-H_{c}^{(i)}\big(s\big)||^2_{\mathcal{H}_2}$ decays monotonically irrespective of the choice of interpolation points and tangential directions. Similarly, a $W$-type CURE also exists; see \cite{panzer2014model} for details.
\subsubsection{FLPORK \cite{zulfiqar2020frequency}}
Let $\mathscr{H}(s)$ and its $r^{th}$-order ROM $\mathscr{H}_r(s)$ be defined as the following
\begin{align}
\mathscr{H}(s)&=C(sI-A)^{-1}B_\Omega,\nonumber\\
\mathscr{H}_r(s)&=C_r(sI-A_r)^{-1}\hat{B}_{\Omega},\nonumber
\end{align} where $B_\Omega=\begin{bmatrix}B &\mathscr{F}(A)B\end{bmatrix}$. Further, let us define $\bar{b}_i$ as $\bar{b}_i=\begin{bmatrix}\mathscr{F}(-\sigma_i)b_i\\b_i\end{bmatrix}$. The input reduction matrix $V_{r,\Omega}$ in FLPORK to enforce the tangential optimality condition $\mathscr{H}(\sigma_i)\bar{b}_i=\mathscr{H}_r(\sigma_i)\bar{b}_i$ is computed as
\begin{align}
AV_{r,\Omega}-V_{r,\Omega}S_r-B_\Omega L_{r,\Omega}=0,\label{666}
\end{align} where $L_{r,\Omega}=\begin{bmatrix}L_r\mathscr{F}(-S_r)\\L_r\end{bmatrix}$. Then $A_r$ is parameterized in $\hat{B}_\Omega$ to place the poles at the mirror images of the interpolation points and to satisfy the following condition
\begin{align}
\mathscr{H}(-\tilde{\lambda}_i)\bar{r}_i=\mathscr{H}_r(-\tilde{\lambda}_i)\bar{r}_i,\label{r2.eq20}
\end{align} where $\bar{r}_i=\begin{bmatrix}\mathscr{F}(\tilde{\lambda}_i)\tilde{r}_i\\\tilde{r}_i\end{bmatrix}$. Let $Q_{s,\Omega}$ be the frequency-limited observability Gramian of the pair $(-S_r,L_r)$ that satisfies the following Lyapunov equation
\begin{align}
-S_r^TQ_{s,\Omega}-Q_{s,\Omega}S_r+\mathscr{F}(-S_r^T)L_r^TL_r+L_r^TL_r\mathscr{F}(-S_r)=0.\nonumber
\end{align}
Then $\mathscr{H}_r(s)$ in FLPORK is computed as
\begin{align}
A_r&=-Q_{s,\Omega}^{-1}S_r^TQ_{s,\Omega},&& \hat{B}_\Omega=\begin{bmatrix}-Q_{s,\Omega}^{-1}L_r^T&-Q_{s,\Omega}^{-1}\mathscr{F}(-S_r^T)L_r^T\end{bmatrix},\nonumber\\
C_r&=CV_{r,\Omega}.\nonumber
\end{align} Finally, $H_r(s)$ that satisfies the optimality condition (\ref{r2.e8}) is computed by setting $B_r=-Q_{s,\Omega}^{-1}L_r^T$. It is noted in \cite{zulfiqar2020frequency} that $Q_{s,\Omega}^{-1}$ is the frequency-limited controllability Gramian $P_{r,\Omega}$ of the pair $(A_r,B_r)=(-Q_{s,\Omega}^{-1}S_r^TQ_{s,\Omega},Q_{s,\Omega}^{-1}L_r^T)$.

Dually a $W$-type FLPORK also exists, which ensures that $H_r(s)$ satisfies the optimality condition (\ref{r2.e7}). Let us define $\mathscr{G}(s)$ and $\mathscr{G}_r(s)$ as the following
\begin{align}
\mathscr{G}(s)&=C_\Omega(sI-A)^{-1}B,\nonumber\\
\mathscr{G}_r(s)&=\hat{C}_\Omega(sI-A_r)^{-1}B_r,\nonumber
\end{align} where $\hat{C}_\Omega=\begin{bmatrix}C\\C\mathscr{F}(A)\end{bmatrix}$. Further, let us define $\bar{c}_i$ as $\bar{c}_i=\begin{bmatrix}\mathscr{F}(-\sigma_i)c_i&c_i\end{bmatrix}$ where $c_i$ are the left tangential directions. The output reduction matrix $W_{r,\Omega}$ in FLPORK to enforce the tangential optimality condition $\bar{c}_i^T\mathscr{H}(\sigma_i)=\bar{c}_i^T\mathscr{H}_r(\sigma_i)$ is computed as
\begin{align}
A^TW_{r,\Omega}-W_{r,\Omega}S_r^T-C_\Omega^TL_{r,\Omega}^T=0,\nonumber
\end{align} where $L_{r,\Omega}=\begin{bmatrix}\mathscr{F}(-S_r)L_r&L_r\end{bmatrix}$. Then $A_r$ is parameterized in $\hat{C}_\Omega$ to place the poles at the mirror images of the interpolation points and to satisfy the following condition
\begin{align}
\bar{l}_i^T\mathscr{G}(-\tilde{\lambda}_i)=\bar{l}_i^T\mathscr{G}_r(-\tilde{\lambda}_i),\label{r2.eq21}
\end{align} where $\bar{l}_i=\begin{bmatrix}\mathscr{F}(\tilde{\lambda}_i)\tilde{l}_i&\tilde{l}_i\end{bmatrix}$. Let $P_{s,\Omega}$ be the frequency-limited controllability Gramian of the pair $(-S_r,L_r)$ that satisfies the following Lyapunov equation
\begin{align}
-S_rP_{s,\Omega}-P_{s,\Omega}S_r^T+\mathscr{F}(-S_r)L_rL_r^T+L_rL_r^T\mathscr{F}(-S_r^T)=0.\nonumber
\end{align}
Then $\mathscr{G}_r(s)$ in FLPORK is computed as
\begin{align}
A_r&=-P_{s,\Omega}S_r^TP_{s,\Omega}^{-1},&B_r&=W_{r,\Omega}^TB,\nonumber\\
\hat{C}_\Omega&=\begin{bmatrix}-L_r^TP_{s,\Omega}^{-1}\\-L_r^T\mathscr{F}(-S_r^T)P_{s,\Omega}^{-1}\end{bmatrix}.\nonumber
\end{align} Finally, $H_r(s)$ that satisfies the optimality condition (\ref{r2.e7}) is computed by setting $C_r=-L_r^TP_{s,\Omega}^{-1}$. It is noted in \cite{zulfiqar2020frequency} that $P_{s,\Omega}^{-1}$ is the frequency-limited observability Gramian $Q_{r,\Omega}$ of the pair $(A_r,C_r)=(-P_{s,\Omega}S_r^TP_{s,\Omega}^{-1},-L_r^TP_{s,\Omega}^{-1})$.
\section{Properties of FLPORK}
There are several properties of FLPORK that were not recognized in the original work. This section fills this gap and presents important properties of FLPORK, some of which can be exploited to formulate an adaptive framework for constructing a ROM. Theorem \ref{th1} gives the equivalence between various optimality conditions and also proves some of the properties of FLPORK.
\begin{theorem}\label{th1}When $H(s)$ and $H_r(s)$ have simple poles, the following statements are true:
\begin{enumerate}
  \item \label{thp1}$\hat{B}_{\Omega}$ and $\hat{C}_{\Omega}$ in FLPORK have particular structures, i.e.,
  \begin{align}
  \hat{B}_{\Omega}&=\begin{bmatrix}B_r&\mathscr{F}(A_r)B_r\end{bmatrix},&\hat{C}_\Omega&=\begin{bmatrix}C_r\\C_r\mathscr{F}(A_r)\end{bmatrix}.\label{r2.e22}
  \end{align}
  \item The optimality condition (\ref{r2.e8}) and the tangential interpolation condition (\ref{r2.eq20}) are equivalent when $\hat{B}_{\Omega}$ has a structure according to (\ref{r2.e22}).
  \item The optimality condition (\ref{r2.e7}) and the tangential interpolation condition (\ref{r2.eq21}) are equivalent when $\hat{C}_{\Omega}$ has a structure according to (\ref{r2.e22}).
  \item The interpolation conditions (\ref{r2.e13}) and (\ref{r2.eq20}) are equivalent when $\hat{B}_{\Omega}$ has a structure according to (\ref{r2.e22}).
  \item \label{thp5}The interpolation conditions (\ref{r2.e12}) and (\ref{r2.eq21}) are equivalent when $\hat{C}_{\Omega}$ has a structure according to (\ref{r2.e22}).
  \item \label{thp6}If the old interpolation data is a subset of the new one, FLPORK ensures that $||H(s)-H_r(s)||_{\mathcal{H}_{2,\Omega}}^2$ decays monotonically.
\end{enumerate}
\end{theorem}
\begin{proof}
\begin{enumerate}
  \item Note that in $V$-type FLPORK, $\mathscr{F}(A_r)=Q_{s,\Omega}^{-1}\mathscr{F}(-S_r^T)Q_{s,\Omega}$ \cite{petersson2013nonlinear} and thus $\mathscr{F}(A_r)Q_{s,\Omega}^{-1}=Q_{s,\Omega}^{-1}\mathscr{F}(-S_r^T)$. Therefore, $-Q_{s,\Omega}^{-1}\mathscr{F}(-S_r^T)L_r^T=\mathscr{F}(A_r)B_r$. Similarly, in $W$-type FLPORK, $\mathscr{F}(A_r)=P_{s,\Omega}\mathscr{F}(-S_r^T)P_{s,\Omega}^{-1}$ and thus $P_{s,\Omega}^{-1}\mathscr{F}(A_r)=\mathscr{F}(-S_r^T)P_{s,\Omega}^{-1}$. Therefore, $-L_r^T\mathscr{F}(-S_r^T)P_{s,\Omega}^{-1}=C_r\mathscr{F}(A_r)$.
      \item  Let the spectral factorization of $A_r$ be $A_r=\tilde{R}\tilde{\Lambda} \tilde{R}^{-1}$ where $\tilde{\Lambda}=diag(\tilde{\lambda}_1,\cdots,\tilde{\lambda}_r)$. Further, define $\tilde{B}$ and $\tilde{C}$ as
\begin{align}
\tilde{B}&=\tilde{R}^{-1}B_r=\begin{bmatrix}\tilde{r}_1&\cdots&\tilde{r}_r\end{bmatrix}^T,&&&\tilde{C}&=C_r\tilde{R}=\begin{bmatrix}\tilde{l}_1&\cdots&\tilde{l}_r\end{bmatrix}.\nonumber
\end{align}
Now define $\hat{\mathscr{P}}_\Omega$ as $\hat{\mathscr{P}}_\Omega=\hat{P}_\Omega\tilde{R}^{-T}$. By noting that $\mathscr{F}(A_r)=\tilde{R}\mathscr{F}(\tilde{\Lambda})\tilde{R}^{-1}$ and multiplying $\tilde{R}^{-T}$ from the right, the equation (\ref{r2.eq.6}) becomes
\begin{align}
A\hat{\mathscr{P}}_\Omega+\hat{\mathscr{P}}_\Omega\tilde{\Lambda}+\mathscr{F}(A)B \tilde{B}^T+B\tilde{B}^T\mathscr{F}(\tilde{\Lambda})=0.\nonumber
\end{align}
Since $\tilde{\Lambda}$ (and resultantly $\mathscr{F}(\tilde{\Lambda})$) is a diagonal matrix \citep{petersson2013nonlinear}, $\hat{\mathscr{P}}_\Omega$ can be computed column-wise as
\begin{align}
\hat{\mathscr{P}}_{\Omega,i}=(-\tilde{\lambda}_iI-A)^{-1}B&\mathscr{F}(\tilde{\lambda}_i)\tilde{r}_i\nonumber\\
&+(-\tilde{\lambda}_iI-A)^{-1}\mathscr{F}(A)B\tilde{r}_i.\nonumber
\end{align}
Further, let us define $\mathscr{P}_{r,\Omega}$ as $\mathscr{P}_{r,\Omega}=P_{r,\Omega}\tilde{R}^{-T}$. By multiplying $\tilde{R}^{-T}$ from the right, the equation (\ref{r2.eq.7}) becomes
\begin{align}
A_r\mathscr{P}_{r,\Omega}+\mathscr{P}_{r,\Omega}\tilde{\Lambda}+\mathscr{F}(A_r)B_r \tilde{B}^T+B_r\tilde{B}^T\mathscr{F}(\tilde{\Lambda})=0.\nonumber
\end{align}
Then $\mathscr{P}_{r,\Omega}$ can be computed column-wise as
\begin{align}
\mathscr{P}_{r,\Omega,i}=(-\tilde{\lambda}_iI-A_r)^{-1}B_r&\mathscr{F}(\tilde{\lambda}_i)\tilde{r_i}\nonumber\\
&+(-\tilde{\lambda}_iI-A_r)^{-1}\mathscr{F}(A_r)B_r\tilde{r_i}.\nonumber
\end{align}
Now, the optimality condition (\ref{r2.e8}) can be written as
\begin{align}
&C\begin{bmatrix}\hat{\mathscr{P}}_{\Omega,1}&\cdots&\hat{\mathscr{P}}_{\Omega,r}\end{bmatrix}\tilde{R}^T\nonumber\\
&\hspace*{2cm}-C_r\begin{bmatrix}\mathscr{P}_{r,\Omega,1}&\cdots&\mathscr{P}_{r,\Omega,r}\end{bmatrix}\tilde{R}^T=0.\label{x1}
\end{align}
After multiplying $\tilde{R}^{-T}$ from the right, each column of (\ref{x1}) becomes
\begin{align}
\mathscr{H}(-\tilde{\lambda}_i)\bar{r}_i-\mathscr{H}_r(-\tilde{\lambda}_i)\bar{r}_i=0.\nonumber
\end{align}
\item Let us define $\hat{\mathscr{Q}}_\Omega$ as $\hat{\mathscr{Q}}_\Omega=\hat{Q}_\Omega\tilde{R}$. Further, define $\tilde{B}$ and $\tilde{C}$ as
\begin{align}
\tilde{B}&=\tilde{R}^{-1}B_r=\begin{bmatrix}\tilde{r}_1&\cdots&\tilde{r}_r\end{bmatrix}^T,&&&\tilde{C}&=C_r\tilde{R}=\begin{bmatrix}\tilde{l}_1&\cdots&\tilde{l}_r\end{bmatrix}.\nonumber
\end{align} By multiplying $\tilde{R}$ from the right, the equation (\ref{r2.eq.8}) becomes
\begin{align}
A^T\hat{\mathscr{Q}}_\Omega+\hat{\mathscr{Q}}_\Omega\tilde{\Lambda}+C^T\tilde{C}\mathscr{F}(\tilde{\Lambda})+\mathscr{F}(A^T)C^T \tilde{C}=0.\nonumber
\end{align}
Then $\hat{\mathscr{Q}}_\Omega$  can be computed column-wise as
\begin{align}
\hat{\mathscr{Q}}_{\Omega,i}=(-\tilde{\lambda}_iI-A^T)^{-1}&C^T\mathscr{F}(\tilde{\lambda}_i)\tilde{l}_i\nonumber\\
&+(-\tilde{\lambda}_iI-A^T)^{-1}\mathscr{F}(A^T)C^T\tilde{l}_i.\nonumber
\end{align}
Now define $\mathscr{Q}_{r,\Omega}$ as $\mathscr{Q}_{r,\Omega}=Q_{r,\Omega}\tilde{R}$. By multiplying $\tilde{R}$ from the right, the equation (\ref{r2.eq.9}) becomes
\begin{align}
A_r^T\mathscr{Q}_{r,\Omega}+\mathscr{Q}_{r,\Omega}\tilde{\Lambda}+C_r^T\tilde{C}\mathscr{F}(\tilde{\Lambda})+\mathscr{F}(A_r^T)C_r^T \tilde{C}=0\nonumber
\end{align}
Then $\mathscr{Q}_{r,\Omega}$  can be computed column-wise as
\begin{align}
\mathscr{Q}_{r,\Omega,i}=(-\tilde{\lambda}_iI-A_r^T)^{-1}&C_r^T\mathscr{F}(\tilde{\lambda}_i)\tilde{l}_i\nonumber\\
&+(-\tilde{\lambda}_iI-A_r^T)^{-1}\mathscr{F}(A_r^T)C_r^T\tilde{l}_i.\nonumber
\end{align}
Further, the optimality condition (\ref{r2.e7}) can be written as
\begin{align}
&B^T\begin{bmatrix}\mathscr{Q}_{r,\Omega,1}&\cdots&\mathscr{Q}_{r,\Omega,r}\end{bmatrix}\tilde{R}\nonumber\\
&\hspace*{2cm}-B_r^T\begin{bmatrix}\mathscr{Q}_{r,\Omega,1}&\cdots&\mathscr{Q}_{r,\Omega,r}\end{bmatrix}\tilde{R}=0.\label{x2}
\end{align}
After multiplying $\tilde{R}^{-1}$ from the right, each column of (\ref{x2}) becomes
\begin{align}
\mathscr{G}^T(-\tilde{\lambda}_i)\bar{l}_i-\mathscr{G}_r^T(-\tilde{\lambda}_i)\bar{l}_i=0.\nonumber
\end{align}
\item  Let the spectral factorization of $A$ be $A=R\Lambda R^{-1}$ where $\Lambda=diag(\lambda_1,\cdots,\lambda_n)$. Note that $\mathscr{F}(A)B=R\mathscr{F}(\Lambda)R^{-1}B$ and $\mathscr{F}(A_r)B_r=\tilde{R}\mathscr{F}(\tilde{\Lambda})\tilde{R}^{-1}B_r$. Thus $\mathscr{H}(s)$ and $\mathscr{H}_r(s)$ can be represented as
\begin{align}
\mathscr{H}(s)&=\begin{bmatrix}\sum_{i=1}^{n}\frac{l_ir_i^T}{s-\lambda_i}&\sum_{i=1}^{n}\frac{l_ir_i^T}{s-\lambda_i}\mathscr{F}(\lambda_i)\end{bmatrix}\nonumber\\
&=\begin{bmatrix}H(s)&H_\Omega(s)\end{bmatrix},\nonumber\\
\mathscr{H}_r(s)&=\begin{bmatrix}\sum_{i=1}^{r}\frac{\tilde{l}_i\tilde{r}_i^T}{s-\tilde{\lambda}_i}&\sum_{i=1}^{r}\frac{\tilde{l}_i\tilde{r}_i^T}{s-\tilde{\lambda}_i}\mathscr{F}(\tilde{\lambda}_i)\end{bmatrix}\nonumber\\
&=\begin{bmatrix}H_r(s)&H_{r,\Omega}(s)\end{bmatrix}.\nonumber
\end{align}
Then it can readily be noted that $\mathscr{H}(-\tilde{\lambda}_i)\bar{r}_i=T_\Omega(-\tilde{\lambda}_i)\tilde{r}_i$ and $\mathscr{H}_r(-\tilde{\lambda}_i)\bar{r}_i=T_{r,\Omega}(-\tilde{\lambda}_i)\tilde{r}_i$.
\item Note that $C\mathscr{F}(A)=CR\mathscr{F}(\Lambda)R^{-1}$ and $C_r\mathscr{F}(A_r)=C_r\tilde{R}\mathscr{F}(\tilde{\Lambda})\tilde{R}^{-1}$. Thus $\mathscr{G}(s)$ and $\mathscr{G}_r(s)$ can be represented as
\begin{align}
\mathscr{G}(s)&=\begin{bmatrix}\sum_{i=1}^{n}\frac{l_ir_i^T}{s-\lambda_i}\\\sum_{i=1}^{n}\frac{l_ir_i^T}{s-\lambda_i}\mathscr{F}(\lambda_i)\end{bmatrix}=\begin{bmatrix}H(s)\\H_\Omega(s)\end{bmatrix},\nonumber\\
\mathscr{G}_r(s)&=\begin{bmatrix}\sum_{i=1}^{r}\frac{\tilde{l}_i\tilde{r}_i^T}{s-\tilde{\lambda}_i}\\\sum_{i=1}^{r}\frac{\tilde{l}_i\tilde{r}_i^T}{s-\tilde{\lambda}_i}\mathscr{F}(\tilde{\lambda}_i)\end{bmatrix}=\begin{bmatrix}H_r(s)\\H_{r,\Omega}(s)\end{bmatrix}.\nonumber
\end{align}
Then it can readily be noted that $\bar{l}_i^T\mathscr{G}(-\tilde{\lambda}_i)=\tilde{l}_i^TT_\Omega(-\tilde{\lambda}_i)$ and $\bar{l}_i^T\mathscr{G}_r(-\tilde{\lambda}_i)=\tilde{l}_i^TT_{r,\Omega}(-\tilde{\lambda}_i)$.
\item From the parts \ref{thp1}-\ref{thp5} of this theorem, it is clear that $V$-type and $W$-type FLPORK ensure the interpolation conditions (\ref{r2.e13}) and (\ref{r2.e12}), respectively. Let an $r_1^{th}$ order ROM $H_{r_1}(s)$ be generated by $V$-type FLPORK by using the interpolation points $(\sigma_1,\sigma_2,\cdots,\sigma_{r_1})$ and the right tangential directions $(b_1,b_2,\cdots,b_{r_1})$. Further, let an $r^{th}$ ($r>r_1$) order ROM $H_{r}(s)$ be generated by $V$-type FLPORK by using the interpolation points $(\sigma_1,\sigma_2,\cdots,\sigma_{r_1},\cdots,\sigma_r)$ and the right tangential directions $(b_1,b_2,\cdots,b_{r_1},\cdots,b_r)$, i.e., the old interpolation data is a subset of the new one. Then it can be readily concluded from the parts \ref{thp1}-\ref{thp5} of this theorem that the following interpolation conditions hold
    \begin{align}
    T_\Omega(-\tilde{\lambda}_i)\tilde{r}_i&=T_{r_1,\Omega}(-\tilde{\lambda}_i)\tilde{r}_i &\textnormal{for}&& i&=1,\cdots,r_1,\nonumber\\
    T_\Omega(-\tilde{\lambda}_i)\tilde{r}_i&=T_{r,\Omega}(-\tilde{\lambda}_i)\tilde{r}_i &\textnormal{for}&& i&=1,\cdots,r,\nonumber\\
    T_{r,\Omega}(-\tilde{\lambda}_i)\tilde{r}_i&=T_{r_1,\Omega}(-\tilde{\lambda}_i)\tilde{r}_i &\textnormal{for}&& i&=1,\cdots,r_1.\nonumber
    \end{align}
    Resultantly, the following holds
    \begin{align}
    ||H(s)-H_{r_1}(s)||_{\mathcal{H}_{2,\Omega}}^2&=||H(s)||_{\mathcal{H}_{2,\Omega}}^2-||H_{r_1}(s)||_{\mathcal{H}_{2,\Omega}}^2,\nonumber\\
    ||H(s)-H_{r}(s)||_{\mathcal{H}_{2,\Omega}}^2&=||H(s)||_{\mathcal{H}_{2,\Omega}}^2-||H_{r}(s)||_{\mathcal{H}_{2,\Omega}}^2,\nonumber\\
    ||H_r(s)-H_{r_1}(s)||_{\mathcal{H}_{2,\Omega}}^2&=||H_r(s)||_{\mathcal{H}_{2,\Omega}}^2-||H_{r_1}(s)||_{\mathcal{H}_{2,\Omega}}^2,\nonumber\\
    ||H_r(s)||_{\mathcal{H}_{2,\Omega}}^2&\geq||H_{r_1}(s)||_{\mathcal{H}_{2,\Omega}}^2,\nonumber\\
    ||H(s)-H_{r}(s)||_{\mathcal{H}_{2,\Omega}}^2&\leq||H(s)-H_{r_1}(s)||_{\mathcal{H}_{2,\Omega}}^2.\label{3000}
    \end{align}
    The monotonic decay property for $W$-type FLPORK can be proved on similar lines by duality, which is left due to brevity.
\end{enumerate}
\end{proof}
\begin{remark}It is interesting to note that the FKIRKA and FLPORK both target the same interpolation conditions, i.e., (\ref{r2.e12}) and (\ref{r2.e13}). This explains why FLIRKA ensures high-fidelity despite being heuristic in nature. However, FLIRKA does not preserve the structure of $\hat{B}_\Omega$ and $\hat{C}_\Omega$ according to (\ref{r2.e22}). Thus it does not satisfy the optimality conditions (\ref{r2.e12}) and (\ref{r2.e13}), in general.\end{remark}

The ROMs generated by PORK and FLPORK have a particular relationship. This is explained in Proposition \ref{prop1}.
\begin{proposition}\label{prop1}
If $S_r$, $L_r$, $Q_{s}$, and $V_r$ are the matrices constructed by $V$-type PORK, then a frequency-limited pseudo-optimal ROM can be constructed from these matrices as
\begin{align}
A_{r,t}&=-S_r^T,&B_{r,t}&=-L_r^T,&C_{r,t}&=CV_{r,\Omega}Q_{s,\Omega}^{-1}\nonumber\end{align} where
\begin{align}
V_{r,\Omega}&=\mathscr{F}(A)V_r+V_r\mathscr{F}(-S_r),\label{e14}\\
Q_{s,\Omega}&=\mathscr{F}(-S_r^T)Q_s+Q_s\mathscr{F}(-S_r).
\end{align}
\end{proposition}
\begin{proof}
We first prove the relationship between $V_r$ and $V_{r,\Omega}$, which are solutions of the Sylvester equations (\ref{6}) and (\ref{666}), respectively. Their relationship can be verified by noting that $S_r\mathscr{F}(-S_r)=\mathscr{F}(-S_r)S_r$ and $A\mathscr{F}(A)=\mathscr{F}(A)A$ \cite{petersson2013nonlinear}, i.e.,
\begin{align}
AV_{r,\Omega}&=A\mathscr{F}(A)V_r+AV_r\mathscr{F}(-S_r)\nonumber\\
AV_{r,\Omega}&=\mathscr{F}(A)AV_r+AV_r\mathscr{F}(-S_r)\nonumber\\
AV_{r,\Omega}&=\mathscr{F}(A)\big[V_rS_r+BL_r\big]+\big[V_rS_r+BL_r\big]\mathscr{F}(-S_r)\nonumber\\
AV_{r,\Omega}&=\mathscr{F}(A)V_rS_r+\mathscr{F}(A)BL_r+\nonumber
\end{align}
\begin{align}
&\hspace*{2cm}V_rS_r\mathscr{F}(-S_r)+BL_r\mathscr{F}(-S_r)\nonumber\\
AV_{r,\Omega}&=\mathscr{F}(A)V_rS_r+\mathscr{F}(A)BL_r+\nonumber
\end{align}
\begin{align}
&\hspace*{2cm}V_r\mathscr{F}(-S_r)S_r+BL_r\mathscr{F}(-S_r)\nonumber\\
AV_{r,\Omega}&=V_{r,\Omega}S_r+B_\Omega L_\Omega.\nonumber
\end{align}
Further, since $Q_{s,\Omega}$ is the frequency-limited observability Gramian of the pair $(-S_r,L_r)$, the following holds (see \cite{gawronski1990model} for details)
\begin{align}
Q_{s,\Omega}&=\mathscr{F}(-S_r^T)Q_s+Q_s\mathscr{F}(-S_r).\nonumber
\end{align}
Moreover, since the pseudo-optimality does not depend on a particular state-space realization, we can apply a similarity transformation on the ROM constructed by PORK using $Q_s^{-1}$ as a transformation matrix, i.e.,
\begin{align}
A_{r,t}=-S_r^T,&& B_{r,t}=-L_r^T,&&C_{r,t}=CV_rQ_s^{-1}.\nonumber
\end{align}
Similarly, by applying a similarity transformation by using $Q_{s,\Omega}^{-1}$ as a transformation matrix, the ROM constructed by FLPORK becomes
\begin{align}
A_{r,t}=-S_r^T,&& B_{r,t}=-L_r^T,&&C_{r,t}=CV_{r,\Omega}Q_{s,\Omega}^{-1}.\nonumber
\end{align}
\end{proof}
\section{Frequency-limited CURE (FLCURE)}
In this section, we aim to develop a recursive framework for FLPORK that enforces new interpolation conditions without affecting the previous ones. This is to exploit the monotonic decay in error property of FLPORK described in the last section. For the infinite frequency case, CURE is essentially such a framework. However, CURE cannot retain the structure of $B_\Omega$ and $C_\Omega$, according to (\ref{r2.e22}). We now generalize CURE for the frequency-limited scenario by using some mathematical manipulations and properties of FLPORK described in the last section.

In CURE, $S_r$, $L_r$, and $V_r$ are always computed recursively as $S_c^{(i)}$, $L_c^{(i)}$, and $V_c^{(i)}$, respectively. If, in addition, PORK is used in every iteration of CURE, $Q_s$ can also be achieved recursively as $Q_c^{(i)}$. If we are able to recursively reproduce $\mathscr{F}(-S_r)$, $Q_{s,\Omega}$, and $V_{r,\Omega }$ from $S_r$, $L_r$, and $V_r$ generated by CURE wherein PORK is used in every iteration, we can recursively generate the frequency-limited pseudo-optimal ROM by using the results of Proposition \ref{prop1}. This is the approach we follow. First, recall that $S_r$, $L_r$, and $V_r$ are recursively computed in CURE as
\begin{align}
S_{c}^{(i)}&=\begin{bmatrix}S_{c}^{(i-1)}&-\bar{B}_{c}^{(i-1)}\tilde{L}^{(i)}\\&\\0&\tilde{S}^{(i)}\end{bmatrix},&&&L_{c}^{(i)}&=\begin{bmatrix}L_{c}^{(i-1)}&&\tilde{L}^{(i)}\end{bmatrix},\nonumber\\
V_{c}^{(i)}&=\begin{bmatrix}V_{c}^{(i-1)}& \tilde{V}^{(i)}\end{bmatrix}.\nonumber
\end{align}
Further, if PORK is used within CURE, $\bar{B}_{c}^{(i)}$ and $Q_{s,c}^{(i)}$ are recursively computed as
\begin{align}
\bar{B}_{c}^{(i)}&=\begin{bmatrix}\bar{B}_c^{(i-1)}\\-(\tilde{Q}_s^{(i)})^{-1}(\tilde{L}^{(i)})^T\end{bmatrix},&&&Q_{s,c}^{(i)}&=\begin{bmatrix}Q_{s,c}^{(i-1)}&0\\0&\tilde{Q}_s^{(i)}\end{bmatrix},\label{r2.eq.31}
\end{align} where
\begin{align}
(-\tilde{S}^{(i)})^T\tilde{Q}_s^{(i)}+\tilde{Q}_s^{(i)}(-S^{(i)})+(\tilde{L}^{(i)})^T(\tilde{L}^{(i)})=0.\label{r2.eq.3222}
\end{align}
Now $\mathscr{F}(-S_c^{(i)})$ can be recursively computed by exploiting the upper triangular structure of $S_c^{(i)}$ and the commutativity property of $\mathscr{F}(-S_c^{(i)})$, i.e., $-S_c^{(i)}\mathscr{F}(-S_c^{(i)})=-\mathscr{F}(-S_c^{(i)})S_c^{(i)}$. The matrix logarithm $\mathscr{F}(-S_c^{(i)})$ is upper triangular since $S_c^{(i)}$ is upper triangular \cite{petersson2013nonlinear}, i.e.,
\begin{align}
\mathscr{F}(-S_{c}^{(i)})=\begin{bmatrix}\mathscr{F}(-S_{c}^{(i-1)})&\mathscr{F}_c^{(i)}\\0&\mathscr{F}(-\tilde{S}^{(i)})\end{bmatrix}.\label{r2.eq.32}
\end{align}
Now we discuss the computation of the off-diagonal entry $\mathscr{F}_c^{(i)}$. Since $-S_c^{(i)}\mathscr{F}(-S_c^{(i)})=-\mathscr{F}(-S_c^{(i)})S_c^{(i)}$, we can write
\begin{align}
&\hspace*{4mm}\begin{bmatrix}S_{c}^{(i-1)}&-\bar{B}_{c}^{(i-1)}\tilde{L}^{(i)}\\&\\0&\tilde{S}^{(i)}\end{bmatrix}\begin{bmatrix}\mathscr{F}(-S_{c}^{(i-1)})&\mathscr{F}_c^{(i)}\\0&\mathscr{F}(-\tilde{S}^{(i)})\end{bmatrix}\nonumber\\
&=\begin{bmatrix}\mathscr{F}(-S_{c}^{(i-1)})&\mathscr{F}_c^{(i)}\\0&\mathscr{F}(-\tilde{S}^{(i)})\end{bmatrix}\begin{bmatrix}S_{c}^{(i-1)}&-\bar{B}_{c}^{(i-1)}\tilde{L}^{(i)}\\&\\0&\tilde{S}^{(i)}\end{bmatrix},\nonumber\\
&\hspace*{4mm}\begin{bmatrix}\star&S_{c}^{(i-1)}\mathscr{F}_c^{(i)}-\bar{B}_{c}^{(i-1)}\tilde{L}^{(i)}\mathscr{F}(-\tilde{S}^{(i)})\\\star&\star\end{bmatrix}\nonumber\\
&=\begin{bmatrix}\star&-\mathscr{F}(-S_{c}^{(i-1)})\bar{B}_{c}^{(i-1)}\tilde{L}^{(i)}+\mathscr{F}_c^{(i)}\tilde{S}^{(i)}\\\star&\star\end{bmatrix}.\nonumber
\end{align}
Thus one can readily note that $\mathscr{F}_c^{(i)}$ can be computed by solving the following Sylvester equation
\begin{align}
&S_c^{(i-1)}\mathscr{F}_c^{(i)}-\mathscr{F}_c^{(i)}\tilde{S}^{(i)}-\bar{B}_c^{(i-1)}\tilde{L}^{(i)}\mathscr{F}(-\tilde{S}^{(i)})\nonumber\\
&\hspace*{3.2cm}+\mathscr{F}(-S_{c}^{(i-1)})\bar{B}_c^{(i-1)}\tilde{L}^{(i)}=0.\label{e23}
\end{align}
We know from Proposition \ref{prop1} that the accumulated reduction subspace $V_{c,\Omega}^{(i)}$ is related to $V_c^{(i)}$ as
\begin{align}
V_{c,\Omega}^{(i)}=\mathscr{F}(A)V_c^{(i)}+V_c^{(i)}\mathscr{F}(-S_c^{(i)}).\label{3666}
\end{align}
By exploiting the upper triangular structure of $\mathscr{F}(-S_c^{(i)})$, $V_{c,\Omega}^{(i)}$ can be obtained recursively as
\begin{align}
V_{c,\Omega}^{(i)}=\begin{bmatrix}V_{c,\Omega}^{(i-1)}&\tilde{V}_{\Omega}^{(i)},\end{bmatrix}
\end{align} where
\begin{align}
\tilde{V}_\Omega^{(i)}&=\mathscr{F}(A)\tilde{V}^{(i)}+\tilde{V}^{(i)}\mathscr{F}(-\tilde{S}^{(i)})+V_c^{(i-1)}\mathscr{F}_c^{(i)}.
\end{align}
Further, we also know from Proposition \ref{prop1} that the accumulated frequency-limited observability Gramian $Q_{s,c,\Omega}^{(i)}$ of the pair $(S_c^{(i)},L_c^{(i)})$ is related to $Q_{s,c}^{(i)}$ as
\begin{align}
Q_{s,c,\Omega}^{(i)}&=\mathscr{F}(-S_{c}^{(i)})^TQ_{s,c}^{(i)}+Q_{s,c}^{(i)}\mathscr{F}(-S_{c}^{(i)}).\label{e22}
\end{align}
By putting (\ref{r2.eq.31}) and (\ref{r2.eq.32}) in (\ref{e22}), one can note that $Q_{s,c,\Omega}^{(i)}$ can be obtained recursively as
\begin{align}
Q_{s,c,\Omega}^{(i)}=\begin{bmatrix}Q_{s,c,\Omega}^{(i-1)}&Q_{s,c}^{(i-1)}\mathscr{F}_c^{(i)}\\(\mathscr{F}_c^{(i)})^TQ_{s,c}^{(i-1)}&\tilde{Q}_{s,\Omega}^{(i)}\end{bmatrix},\label{4111}
\end{align}
where
\begin{align}
\tilde{Q}_{s,\Omega}^{(i)}=\mathscr{F}(-\tilde{S}^{(i)})^T\tilde{Q}_s^{(i)}+\tilde{Q}_s^{(i)}\mathscr{F}(-\tilde{S}^{(i)}).
\end{align}
Finally, the accumulated frequency-limited pseudo-optimal ROM can be obtained as
\begin{align}
A_c^{(i)}&=-(S_c^{(i)})^T,&& B_c^{(i)}&=-(L_c^{(i)})^T,&& C_c^{(i)}&=CV_{c,\Omega}^{(i)}(Q_{s,c,\Omega}^{(i)})^{-1}.\nonumber
\end{align}
Since $H_{c}^{(i)}\big(s\big)$ remains a frequency-limited pseudo-optimal ROM as $i$ grows, and the old interpolation data $(S_c^{(i-1)},L_c^{(i-1)})$ remains embedded in the new interpolation data $(S_c^{(i)},L_c^{(i)})$, the error $||H(s)-H_{c}^{(i)}\big(s\big)||^2_{\mathcal{H}_{2,\Omega}}$ decays monotonically as proved in the part \ref{thp6} of Theorem \ref{th1}. Moreover, since the ROM is frequency-limited pseudo-optimal, the following holds
\begin{align}
&||H(s)-H_{c}^{(i)}\big(s\big)||^2_{\mathcal{H}_{2,\Omega}}\nonumber\\
&\hspace*{2.5cm}=CP_\Omega C^T-CV_{c,\Omega}^{(i)}(Q_{s,c,\Omega}^{(i)})^{-1}(V_{c,\Omega}^{(i)})^TC^T\nonumber\\
&\hspace*{2.5cm}=C\Big(P_\Omega-V_{c,\Omega}^{(i)}(Q_{s,c,\Omega}^{(i)})^{-1}(V_{c,\Omega}^{(i)})^T\Big)C^T.\nonumber
\end{align} Thus $\hat{P}_\Omega=V_{c,\Omega}^{(i)}(Q_{s,c,\Omega}^{(i)})^{-1}(V_{c,\Omega}^{(i)})^T$ monotonically approaches $P_\Omega$ after each iteration. Hence, FLCURE also provides the approximation of $P_\Omega$. The pseudo-code of $V$-type FLCURE is presented in Algorithm \ref{Alg1}.
\begin{algorithm}[!h]
  \begin{algorithmic}[1]
      \State Initialize $B_{\perp}^{(0)}=B$, $S_{c}^{(0)}=[\hspace{2mm}]$, $C_{c}^{(0)}=[\hspace{2mm}]$, $L_{c}^{(0)}=[\hspace{2mm}]$, $Q_{s,c}^{(0)}=[\hspace{2mm}]$, $Q_{s,c,\Omega}^{(0)}=[\hspace{2mm}]$, $\bar{B}_{c}^{(0)}=[\hspace{2mm}]$, $V_{c}^{(0)}=[\hspace{2mm}]$, $V_{c,\Omega}^{(0)}=[\hspace{2mm}]$.
      \State \textbf{for} $i=1,\cdots,k$
      \State $Ran(\tilde{V}^{(i)})=\underset {j=1,\cdots,r_j}{span}\{(\sigma_jI-A)^{-1}Bb_j\}$.
      \State Set $\tilde{W}^{(i)}=\tilde{V}^{(i)}((\tilde{V}^{(i)})^T\tilde{V}^{(i)})^{-1}$, $\tilde{A}=(\tilde{W}^{(i)})^TA\tilde{V}^{(i)}$, $\tilde{B}=(\tilde{W}^{(i)})^TB_\bot^{(i-1)}$, $B_\bot^{(i)}=B_\bot^{i-1}-\tilde{V}^{(i)}\tilde{B}$, $\tilde{L}^{(i)}=\big((B_\bot^{(i)})^TB_\bot^{(i)}\big)^{-1}(B_\bot^{(i)})^T(A\tilde{V}^{(i)}-\tilde{V}^{(i)}\tilde{A})$, $\tilde{S}^{(i)}=\tilde{A}-\tilde{B}\tilde{L}^{(i)}$.
      \State Solve the equation (\ref{r2.eq.3222}) to compute $\tilde{Q}_s^{(i)}$.
      \State $Q_{s,c}^{(i)}=\begin{bmatrix}Q_{s,c}^{(i-1)}&0\\0&\tilde{Q}_s^{(i)}\end{bmatrix}$, $\bar{B}_{c}^{(i)}=\begin{bmatrix}\bar{B}_{c}^{(i-1)}\\-(\tilde{Q}_s^{(i)})^{-1}\big(\tilde{L}^{(i)}\big)^T\end{bmatrix}$, $S_{c}^{(i)}=\begin{bmatrix}S_{c}^{(i-1)}&-\bar{B}_{c}^{(i-1)}\tilde{L}^{(i)}\\0&\tilde{S}^{(i)}\end{bmatrix}$, $L_{c}^{(i)}=\begin{bmatrix}L_{c}^{(i-1)}&\tilde{L}^{(i)}\end{bmatrix}$, $V_{c}^{(i)}=\begin{bmatrix}V_{c}^{(i-1)}& \tilde{V}^{(i)}\end{bmatrix}$.
      \State Solve the equation (\ref{e23}) to compute $\mathscr{F}_c^{(i)}$.
      \State $\mathscr{F}(-S_{c}^{(i)})=\begin{bmatrix}\mathscr{F}(-S_{c}^{(i-1)})&\mathscr{F}_c^{(i)}\\0&\mathscr{F}(-\tilde{S}^{(i)})\end{bmatrix}$.
      \State $\tilde{V}_\Omega^{(i)}=\mathscr{F}(A)\tilde{V}^{(i)}+\tilde{V}^{(i)}\mathscr{F}(-\tilde{S}^{(i)})+V_c^{(i-1)}\mathscr{F}_c^{(i)}$.
      \State $\tilde{Q}_{s,\Omega}^{(i)}=\mathscr{F}(-\tilde{S}^{(i)})^T\tilde{Q}_s^{(i)}+\tilde{Q}_s^{(i)}\mathscr{F}(-\tilde{S}^{(i)})$.
      \State $V_{c,\Omega}^{(i)}=\begin{bmatrix}V_{c,\Omega}^{(i-1)}&\tilde{V}_{\Omega}^{(i)}\end{bmatrix}$.
      \State $Q_{s,c,\Omega}^{(i)}=\begin{bmatrix}Q_{s,c,\Omega}^{(i-1)}&Q_{s,c}^{(i-1)}\mathscr{F}_c^{(i)}\\(\mathscr{F}_c^{(i)})^TQ_{s,c}^{(i-1)}&\tilde{Q}_{s,\Omega}^{(i)}\end{bmatrix}$.
    \State $A_c^{(i)}=-(S_c^{(i)})^T$, $B_c^{(i)}=-(L_c^{(i)})^T$, $C_c^{(i)}=CV_{c,\Omega}^{(i)}(Q_{s,c,\Omega}^{(i)})^{-1}$.
  \end{algorithmic}
  \caption{FLCURE (V-type)}\label{Alg1}
\end{algorithm}

We now present a $W$-type algorithm for FLCURE without proof for brevity. The pseudo-code of FLCURE (W-type) is presented in Algorithm \ref{Alg2}. The W-type algorithm for FLCURE also ensures that $||H(s)-H_{c}^{(i)}\big(s\big)||^2_{\mathcal{H}_{2,\Omega}}$ decays monotonically irrespective of the choice of interpolation points and the tangential directions as $i$ grows.

Note that
\begin{align}
&||H(s)-H_{c}^{(i)}\big(s\big)||^2_{\mathcal{H}_{2,\Omega}}\nonumber\\
&\hspace*{2.2cm}=B^TQ_\Omega B-B^TW_{c,\Omega}^{(i)}(P_{s,c,\Omega}^{(i)})^{-1}(W_{c,\Omega}^{(i)})^TB\nonumber\\
&\hspace*{2.2cm}=B^T\Big(Q_\Omega-W_{c,\Omega}^{(i)}(P_{s,c,\Omega}^{(i)})^{-1}(W_{c,\Omega}^{(i)})^T\Big)B.\nonumber
\end{align}Therefore, $\hat{Q}_\Omega=W_{c,\Omega}^{(i)}(P_{s,c,\Omega}^{(i)})^{-1}(W_{c,\Omega}^{(i)})^T$ monotonically approaches $Q_\Omega$ after each iteration of FLCURE. Hence, it also provides an approximation of $Q_\Omega$.
\begin{algorithm}[!h]
  \begin{algorithmic}[1]
      \State Initialize $C_{\perp}^{(0)}=C$, $S_{c}^{(0)}=[\hspace{2mm}]$, $C_{c}^{(0)}=[\hspace{2mm}]$, $L_{c}^{(0)}=[\hspace{2mm}]$, $P_{s,c}^{(0)}=[\hspace{2mm}]$, $P_{s,c,\Omega}^{(0)}=[\hspace{2mm}]$, $\bar{C}_{c}^{(0)}=[\hspace{2mm}]$, $W_{c}^{(0)}=[\hspace{2mm}]$, $W_{c,\Omega}^{(0)}=[\hspace{2mm}]$.
      \State \textbf{for} $i=1,\cdots,k$
\State $Ran(\tilde{W}^{(i)})=\underset {j=1,\cdots,r_j}{span}\{(\sigma_jI-A^T)^{-1}C^Tc_j^T\}$.
\State Set $\tilde{V}^{(i)}=\tilde{W}^{(i)}$, $\tilde{W}^{(i)}=\tilde{W}^{(i)}\big((\tilde{W}^{(i)})^T\tilde{W}^{(i)}\big)^{-1}$, $\tilde{A}=(\tilde{W}^{(i)})^TA\tilde{V}^{(i)}$, $\tilde{C}=C_\bot^{(i-1)}\tilde{V}^{(i)}$, $C_\bot^{(i)}=C_\bot^{i-1}-\tilde{C}(\tilde{W}^{(i)})^T$, $\tilde{L}^{(i)}=\big((\tilde{W}^{(i)})^TA-\tilde{A}(\tilde{W}^{(i)})^T\big)(C_\bot^{(i)})^T\big(C_\bot^{(i)}(C_\bot^{(i)})^T\big)^{-1}$, $\tilde{S}^{(i)}=\tilde{A}-\tilde{L}^{(i)}\tilde{C}$.
\State Solve $-\tilde{S}^{(i)}\tilde{P}_s^{(i)}-\tilde{P}_s^{(i)}(\tilde{S}^{(i)})^T+\tilde{L}^{(i)}(\tilde{L}^{(i)})^T=0$.
\State $P_{s,c}^{(i)}=\begin{bmatrix}P_{s,c}^{(i-1)}&0\\0&\tilde{P}_s^{(i)}\end{bmatrix}$, $\bar{C}_{c}^{(i)}=\begin{bmatrix}\bar{C}_{c}^{(i-1)}\\-(\tilde{L}^{(i)})^T(\tilde{P}_s^{(i)})^{-1}\end{bmatrix}$, $S_{c}^{(i)}=\begin{bmatrix}S_{c}^{(i-1)}&0\\-\tilde{L}^{(i)}\bar{C}_{c}^{(i-1)}&\tilde{S}^{(i)}\end{bmatrix}$, $L_{c}^{(i)}=\begin{bmatrix}L_{c}^{(i-1)}\\\tilde{L}^{(i)}\end{bmatrix}$, $W_{c}^{(i)}=\begin{bmatrix}W_{c}^{(i-1)}& \tilde{W}^{(i)}\end{bmatrix}$.
\State Solve $-\tilde{S}^{(i)}\mathscr{F}_c^{(i)}+\mathscr{F}_c^{(i)}S_c^{(i-1)}-\mathscr{F}(-\tilde{S}^{(i)})\tilde{L}^{(i)}\bar{C}_{c}^{(i-1)}+\tilde{L}^{(i)}\bar{C}_{c}^{(i-1)}\mathscr{F}(-S_c^{(i-1)})=0.$
\State $\mathscr{F}(-S_c^{(i)})=\begin{bmatrix}\mathscr{F}(-S_c^{(i-1)})& 0\\\mathscr{F}_c^{(i)}&\mathscr{F}(-\tilde{S}^{(i)})\end{bmatrix}$.
\State $\tilde{W}_\Omega^{(i)}=\mathscr{F}(A)^T\tilde{W}^{(i)}+\tilde{W}^{(i)}\mathscr{F}(-\tilde{S}^{(i)})^T+W_{c}^{(i-1)}(\mathscr{F}_c^{(i)})^T$.
\State $\tilde{P}_{s,\Omega}^{(i)}=\mathscr{F}(\tilde{S}^{(i)})\tilde{P}_s^{(i)}+\tilde{P}_s^{(i)}\mathscr{F}(\tilde{S}^{(i)})^T$.
\State $P_{s,c,\Omega}^{(i)}=\begin{bmatrix}P_{s,c,\Omega}^{(i-1)}&Q_{s,c}^{(i-1)}(\mathscr{F}_c^{(i)})^T\\
      \mathscr{F}_c^{(i)}Q_{s,c}^{(i-1)}&\tilde{P}_{s,\Omega}^{(i)}\end{bmatrix}.$
\State $W_{c,\Omega}^{(i)}=\begin{bmatrix}W_{c,\Omega}^{(i-1)}&\tilde{W}_\Omega^{(i)}\end{bmatrix}$.
\State $A_{c}^{(i)}=(-S_{c}^{(i)})^T$, $B_{c}^{(i)}=(P_{s,c,\Omega}^{(i)})^{-1}(W_{c,\Omega}^{(i)})^TB$, $C_{c}^{(i)}=(-L_{c}^{(i)})^T$.
\State \textbf{end for}
  \end{algorithmic}
  \caption{FLCURE (W-type)}\label{Alg2}
\end{algorithm}
\begin{remark}
The computation of $\mathscr{F}(A)$ is expensive in a large-scale setting; however, an efficient Krylov subspace-based approximation can be obtained by using the procedure described in \cite{benner2016frequency}. Note that the Krylov subspace required for this purpose is not needed to be computed exclusively as it is already available in Algorithms 1 and 2, i.e., $V_c^{(i)}$ and $W_c^{(i)}$.
\end{remark}
\begin{remark} The poles of the ROM in FLPORK and FLCURE are the mirror images of the interpolation points. Since the interpolation points are always selected in the right-half of the $s$-plane, the ROM is guaranteed to be stable.
\end{remark}
\begin{remark} A frequency-limited pseudo-optimal ROM generated by a single run of FLPORK and that by the multiple steps of FLCURE are equivalent if the same interpolation points and tangential directions are used. The significance of FLCURE and its monotonic decay in the error property is that an adaptive choice of the order $r$ of the ROM can be made, i.e., the order of the ROM can be increased until the $\mathcal{H}_{2,\Omega}$-norm error decays below the desired tolerance.
\end{remark}
\subsection{Connection with the ADI Method}
We have seen, so far, that to generalize any infinite frequency-interval moment matching method to the frequency-limited scenario, the only information that is required is the matrices of the Sylvester equations (\ref{666}), i.e., $S_r$ and $L_r$. Interestingly, the Krylov subspace-based methods and LR-ADI methods for the solution of large-scale Lyapunov equation are shown equivalent in \cite{wolf2014h}. It is further shown in \cite{wolf2014h} that the LR-ADI method implicitly performs $\mathcal{H}_2$-pseudo-optimal MOR. Moreover, the matrices $S_r$ and $L_r$ matrices of their respective Sylvester equations are also derived, which are shown to be equivalent to those used in PORK. We now briefly discuss how LR-ADI method can effortlessly to be generalized to the frequency-limited scenario by using the results from the last subsection.

Let $\hat{P}$ be the approximation of the controllability Gramian $P$ of the pair $(A,B)$. It is shown in \cite{li2002low} and \cite{penzl1999cyclic} that the low-rank Cholesky factor $\hat{Z}$ of $\hat{P}$, i.e., $\hat{P}=\hat{Z}\hat{Z}^*$, can be computed using ADI iterations as the following
\begin{align}
\hat{Z}_1&=\sqrt{2 Re(\sigma_1)}\big(A-\sigma_1 I\big)^{-1}B,\nonumber\\
\hat{Z}_j&=\sqrt{\frac{Re(\sigma_j)}{Re(\sigma_{j-1})}}\Big(I+(\sigma_j+\bar{\sigma}_{j-1})\big(A-\sigma_j I\big)^{-1}\Big)\hat{Z}_{j-1},\nonumber\\
\hat{Z}&=\begin{bmatrix}\hat{Z}_1,\cdots,\hat{Z}_k\end{bmatrix},\nonumber
\end{align} for $j=1,\cdots,k$. It is further shown in \cite{wolf2014h} that $\hat{P}$ constructed using this method is equivalent to the one generated by CURE when PORK is used in each iteration, i.e., $\hat{P}=V_c^{(i)}(Q_{s,c}^{(i)})^{-1}\big(V_c^{(i)}\big)^T$. Let $\alpha_j=\sqrt{2 Re(\sigma_j)}$, then $\hat{Z}$ satisfies the following Sylvester equation
\begin{align}
A\hat{Z}-\hat{Z}S_c^{(i)}-BL_c^{(i)}=0\nonumber
\end{align} where
\begin{align}
S_c^{(i)}&=\begin{bmatrix}\sigma_1I_m& \alpha_1\alpha_2I_m&\cdots&\alpha_1\alpha_kI_m\\
&\ddots&\ddots&\vdots\\
&&\ddots&\alpha_{k-1}\alpha_kI_m\\
&&&\sigma_kI_m\end{bmatrix},\nonumber\\
L_c^{(i)}&=\begin{bmatrix}\alpha_1I_m,\cdots,\alpha_kI_m\end{bmatrix},\nonumber
\end{align} $Q_{s,c}^{(i)}=I$, $V_c^{(i)}=\hat{Z}$, and $I_m$ is an $m\times m$ identity matrix. Using the relations (\ref{3666}) and (\ref{e22}), it can readily be concluded that $\hat{P}_\Omega=V_{c,\Omega}^{(i)}(Q_{s,c,\Omega}^{(i)})^{-1}\big(V_{c,\Omega}^{(i)}\big)^T$ obtained in the last subsection is related to the LR-ADI method. In fact, the frequency-limited generalization of the LR-ADI method already exists and can be found with a detailed discussion in \cite{benner2016frequency}. Thus the results of this subsection show that the frequency-limited LR-ADI method of Lyapunov equations implicitly performs frequency-limited $\mathcal{H}_2$-pseudo-optimal MOR. This is consistent with the parallelism between the infinite- and finite-interval methods.

It is evident from ($I_m$ and) the structure of $S_c^{(i)}$ and $L_c^{(i)}$ that $\hat{Z}$ corresponds to a block-Krylov subspace \cite{wolf2014h}. To obtain additional freedom of tangential directions, the connection between the Krylov subspace and the LR-ADI method can be generalized to incorporate tangential directions. This is the central theme of the algorithm proposed in \cite{wolf2016adi}. Moreover, it is shown in \cite{wolf2014h} that this algorithm is equivalent to the Krylov subspace-based restart scheme proposed in \cite{ahmad2010krylov}. Again, by using the relations (\ref{3666}) and (\ref{e22}), these algorithms can similarly be generalized to the frequency-limited scenario. Note that Algorithms 1 and 2 already have the freedom to incorporate tangential directions. In short, though simple in their formulation, the relations (\ref{3666}) and (\ref{e22}) can generalize a whole family of algorithms for the solution of Lyapunov equations and MOR to the frequency-limited case.
\subsection{Adaptive Selection of Order}
To adaptively decide the order of the ROM, FLCURE needs a stopping criterion. The computation of $||H(s)-H_{c}^{(i)}(s)||_{\mathcal{H}_{2,\Omega}}=\sqrt{C\big(P_\Omega-\hat{P}_{\Omega}\big)C^T}$ is computationally expensive in a large-scale setting and thus not feasible as a stopping criterion. FLCURE does provide an approximation of $P_\Omega$; however, the difference $P_\Omega-\hat{P}_\Omega$ corresponds to $||H(s)-H_{c}^{(i)}(s)||_{\mathcal{H}_{2,\Omega}}$ itself, and thus $\hat{P}_\Omega$ generated by FLCURE cannot be used as an error estimator. One possible approach can be to run a few iterations of FLIRKA and generate a ROM of much greater size than we are after (so that it contains most of the dynamics of the original system) and compute an estimate of $P_\Omega$ as $\bar{P}_\Omega=V_{r,\Omega}P_{r,\Omega}V_{r,\Omega}^T$. The interpolation data set generated by FLIRKA can later be used for an appropriate selection of interpolation points in FLCURE, which we will discuss in the next subsection. Then $\bar{P}_\Omega$ can be used to compute an estimate of error $||H(s)-H_{c}^{(i)}(s)||_{\mathcal{H}_{2,\Omega}}$, and FLCURE can increase the error of the ROM until the error is less than the desired tolerance. A conservative value of tolerance can compensate for the approximation used in the error estimation.

In case the performance specification is not provided by the user in terms of the desired error tolerance, the following methodology can be adopted to decide the final approximation quality of the ROM on its own. The advantage of using FLCURE is that the ROM continues to satisfy (\ref{r2.e10}) as the algorithm progresses. Thus the approximation error expression only depends on $||H_{c}^{(i)}(s)||^2_{\mathcal{H}_{2,\Omega}}$, i.e., the decay in approximation error is encoded in the rise of $\mathcal{H}_{2,\Omega}$-norm of the accumulated ROM. Therefore, we use the relative rate of the rise in $||H_{c}^{(i)}(s)||^2_{\mathcal{H}_{2,\Omega}}$ as the stopping criteria of FLCURE. Now define $\epsilon_c^{(i)}$ as
\begin{align}
\epsilon_c^{(i)}=\frac{||\big(H_{c}^{(i)}(s)-H_{c}^{(i-1)}(s)\big)||_{\mathcal{H}_{2,\Omega}}}{||\big(H_{c}^{(i-1)}(s)||_{\mathcal{H}_{2,\Omega}}}.\nonumber
\end{align}
Also define the following slope
\begin{align}
\epsilon_{slope}^{(i)}=\frac{\epsilon_c^{(i)}-\epsilon_c^{(i-1)}}{r_i-r_{i-1}}\nonumber
\end{align} where $r_i$ and $r_{i-1}$ are the orders of $H_{c}^{(i)}(s)$ and $H_{c}^{(i-1)}(s)$, respectively. Note that the computation of $\epsilon_{slope}^{(i)}$ only involves small-scale operations, and hence its computation is feasible even in a large-scale setting. $\epsilon_{slope}^{(i)}$ is a quantitative measure of how quickly the approximation error is decaying in FLCURE. We suggest stopping FLCURE if $\epsilon_{slope}^{(i)}$ remains below the desired tolerance for several consecutive iterations because this shows that the approximation accuracy is not appreciably improving with the increase in the order of the ROM. Additionally, constraints like maximum allowable order or number of iterations can be used to terminate the algorithm prematurely, even if $\epsilon_{slope}^{(i)}$ keeps on rising above the desired rate. It should be stressed here that without the property (\ref{3000}) of FLCURE, the expectation that $\epsilon_{c}^{(i)}$ keeps on rising with the increase in the order of the ROM would have been merely a hope.

If FLCURE is used to obtain an approximation of $P_\Omega$, the following stopping criterion can be used to adaptively terminate the algorithm, i.e., when
\begin{align}
\epsilon_{rel}^{(i)}=\frac{||A\hat{P}_\Omega+\hat{P}_\Omega A^T+\mathscr{F}(A)BB^T+BB^T\mathscr{F}(A)^T||}{||\mathscr{F}(A_c^{i})B_c^{(i)}(B_c^{(i)})^T+B_c^{(i)}(B_c^{(i)})^T(\mathscr{F}(A_c^{i}))^T||}\nonumber
\end{align}
is less than the desired tolerance. As shown in \cite{benner2016frequency} that this scaled norm of the Lyapunov residual matrix can be efficiently computed without dealing with $A$, $B$, $\mathscr{F}(A)$, and $\hat{P}_\Omega$. Similarly, if FLCURE is used to obtain an approximation of $Q_\Omega$, the following stopping criterion can be used to adaptively terminate the algorithm, i.e., when
\begin{align}
\epsilon_{rel}^{(i)}=
\frac{||A^T\hat{Q}_\Omega+\hat{Q}_\Omega A+\mathscr{F}(A)^TC^TC+C^TC\mathscr{F}(A)||}{||\mathscr{F}(A_c^{i})^T(C_c^{(i)})^TC_c^{(i)}+(C_c^{(i)})^TC_c^{(i)}\mathscr{F}(A_c^{i})||}\nonumber
\end{align}
is less than the desired tolerance. Additionally, constraints like maximum allowable iterations and unacceptable rate of decay in the scaled norm of the Lyapunov residual matrix can be used to terminate the algorithm prematurely.
\subsection{Adaptive Selection of Interpolation Data}
We have, so far, established a theoretical guarantee on the error that it will continue to decay as the number of iterations of FLCURE grows. However, it says nothing about how quickly and steeply the error decays, upon which the stopping criteria rely. This depends on the choice of interpolation data. The selection of interpolation data is one of the major drawbacks of Krylov subspace-based MOR algorithms as there is no straightforward choice that guarantees supreme accuracy. There exist some iterative schemes like IRKA \cite{gugercin2008h_2} that result in good accuracy, even if they are initialized with fairly random interpolation data. Moreover, some practical guidelines are also reported in the literature to ensure good accuracy. We divide the choices of interpolation data into two cases, i.e., (i) Heuristic selection (ii) Optimization-based selection. The optimization of interpolation data is a topic in its own right; nevertheless, we describe a possible sketch of an optimization scheme.
\subsubsection{Heuristic Selection}
In many practical situations, like in power systems \cite{zulfiqar2019finite,zulfiqar2020frequency}, vibratory systems \cite{rook1996modal}, and lossless systems \cite{konkel2014posteriori}, the preservation of particular poles in the ROM is important from a physical standpoint. FLPORK (and resultantly FLCURE) has the pole-placement property, i.e., the poles of the ROM are the mirror images of the interpolation points specified by the user. Thus the mirror images of the poles to be preserved are the obvious choice in these situations. Moreover, the peaks and dips in the frequency response of the linear systems correspond to the poles with large residuals. Further, it is shown in \cite{gugercin2008h_2}, the interpolation at the mirror images of these poles (with their respective residuals used as tangential directions) ensures a small $\mathcal{H}_2$-norm error. These modes can efficiently be captured in a large-scale setting using the eigensolver proposed in \cite{rommes2006efficient}. If a ROM of appropriate accuracy is successfully achieved by using this interpolation data alone, then such a ROM maintains a physical link with the original system by keeping its dominant poles. If, however, this interpolation data is insufficient for achieving the desired accuracy, then IRKA and FLIRKA can be used to generate a good interpolation data set wherein IRKA is initialized randomly, and the final interpolation data is used to initialize FLIRKA. This strategy is tested in \cite{vuillemin2013h2,zulfiqar2020frequency}, and some promising results are achieved. Both IRKA and FLIRKA can be stopped prematurely after a few iterations, as we are not interested in constructing the ROM using these algorithms. The interpolation points selected must lie in the right half of the $s$-plane. As discussed earlier that their ROM can also be used to compute an error estimator based on which an automatic selection of order can be made. We will show in the next section via numerical simulations that this strategy is an effective one that yields sufficient accuracy. Therefore, this selection criteria can be used to remove the dependence on the user inference.
\subsubsection{Optimization-based Selection}
We now discuss a basic sketch of an optimization-based scheme for the selection of interpolation data if an optimization toolbox is at our disposal. Now suppose that $H(s)$ is a SISO system, an interim ROM of order $2$ is constructed at each iteration of FLCURE, and the final ROM is the accumulation of all these interim ROMs. As discussed in the last subsection, the cost function for minimizing error in FLCURE depends only on the transfer function of the ROM due to the property (\ref{r2.e10}), i.e.,
\begin{align}
\mathscr{J}=-||H_c^{(i)}(s)||^2_{\mathcal{H}_{2,\Omega}}=-C_c^{i}(Q_{s,c,\Omega}^{(i)})^{-1}(C_c^{(i)})^T.\nonumber
\end{align}
Since we want to judiciously enforce frequency-limited pseudo-optimality on a $\mathcal{H}_2$-pseudo-optimal model at each iteration of CURE, we borrow such a ROM from \cite{panzer2014model}, i.e.,
\begin{align}
\tilde{A}^{(i)}=-(\tilde{S}^{(i)})^T,\hspace*{1mm}\tilde{B}^{(i)}=-(\tilde{L}^{(i)})^T,\hspace*{1mm}\tilde{C}^{(i)}=C\tilde{V}^{(i)}(\tilde{Q}_s^{(i)})^{-1},\nonumber
\end{align} where
\begin{align}
\tilde{S}^{(i)}&=\begin{bmatrix}2a&\sqrt{b}\\-\sqrt{b}&0\end{bmatrix},\hspace*{1mm}\tilde{L}^{(i)}=\begin{bmatrix}1&0\end{bmatrix},\hspace*{1mm}\tilde{Q}_s^{(i)}=\begin{bmatrix}\frac{1}{4a}&0\\0&\frac{1}{4a}\end{bmatrix},\nonumber\\
\tilde{V}&=\begin{bmatrix}A_{\sigma_2}^{-1}AA_{\sigma_1}^{-1}B&&\sqrt{b}A_{\sigma_1}^{-1}A_{\sigma_1}^{-1}B\end{bmatrix},\nonumber\\
A_{\sigma_1}&=A-\sigma_1I,\hspace*{1mm}A_{\sigma_2}=A-\sigma_2I,\hspace*{1mm}\sigma_{1,2}=a\pm\sqrt{a^2-b},\nonumber
\end{align}$a$ and $b$ are positive real numbers, and the two poles of the interim ROM are $-\sigma_{1,2}$. It can easily be verified that this interim ROM is always stable. $V_{c,\Omega}^{(i)}$ and $Q_{s,c,\Omega}^{(i)}$ can be recursively updated according to (\ref{3666})-(\ref{4111}) by using $\tilde{V}$ and $\tilde{Q}_s^{(i)}$, respectively, and $C_c^{(i)}$ can be updated as $C_c^{(i)}=CV_{c,\Omega}^{(i)}(Q_{s,c,\Omega}^{(i)})^{-1}$ at each iteration. The constrained minimization problem under consideration only has two free parameters, $a$ and $b$, i.e.,
\begin{equation*}
\begin{aligned}
&\underset{a>0,b>0}{\text{minimize}}
& & \mathscr{J}(a,b).
\end{aligned}
\end{equation*} This optimization problem can be solved by MATLAB's ``\textit{fmincon}" command. To speed up the optimization process, the analytic gradient and Hessian matrices can be supplied to the optimization toolbox. Since it involves only two parameters, these can be supplied even by MATLAB's symbolic toolbox. To extend this optimization scheme to MIMO systems, one can use the subsystem-based approach used in \cite{jiang2018model} to generalize the cross Gramian based-MOR to MIMO systems; see \cite{jiang2018model} for more details. Let us divide $H(s)$ into $p\times m$ subsystems by partitioning $B$ and $C$ as
\begin{align}
B=\begin{bmatrix}B_1&\cdots&B_m\end{bmatrix}\textnormal{ and }C=\begin{bmatrix}C_1^T&\cdots&C_p^T\end{bmatrix}^T.\nonumber
\end{align} Then each subsystem of $H(s)$ becomes $H_{xy}(s)=C_x(sI-A)^{-1}B_y$. Each subsystem can be reduced to obtain $p\times m$ ROMs, i.e., $\tilde{H}_{xy}(s)=\tilde{C}_{xy}(sI-\tilde{A}_{xy})^{-1}\tilde{B}_{xy}$. These ROMs can then be packed to obtain the final ROM $(A_r,B_r,C_r)$ as the following
\begin{align}
A_r&=\oplus_{x=1}^p \hat{A}_{x},\hspace*{2mm}B_r=\begin{bmatrix}\hat{B}_1^T&\cdots&\hat{B}_m^T\end{bmatrix}^T,\hspace*{2mm}C_r=\oplus_{x=1}^p\hat{C}_x,\nonumber\\
\hat{A}_x&=\oplus_{y=1}^m\tilde{A}_{xy},\hspace*{2mm}\hat{B}_x=\oplus_{y=1}^m\tilde{B}_{xy},\hspace*{2mm}\hat{C}_x=\begin{bmatrix}C_{x1}&\cdots&C_{xm}\end{bmatrix}.\nonumber
\end{align}
Unfortunately, the final ROM $(A_r,B_r,C_r)$ obtained with this approach is no more frequency-limited pseudo-optimal, unlike the subsystems $\tilde{H}_{xy}(s)=\tilde{C}_{xy}(sI-\tilde{A}_{xy})^{-1}\tilde{B}_{xy}$. However, the ROM remains guaranteed to be stable as each subsystem generated by FLCURE is stable \cite{jiang2018model}.
\subsection{Computational Cost}
In the subsection, we analyze and compare the computational costs of FLBT, FLIRKA, FLPORK, and FLCURE. The most expensive operation (i.e., of the order $\mathcal{O}(n^2)$) in FLBT is the solution of Lyapunov equations (\ref{e2}) and (\ref{e3}) that makes it infeasible for large-scale systems. FLIRKA, FLPORK, and FLCURE avoid the computation of these Lyapunov equations, and their main computational effort is the Sylvester equation (\ref{666}). FLIRKA has two such Sylvester equations, i.e., $V_{r,\Omega}$ and $W_{r,\Omega}$, and it is an iterative algorithm that solves these Sylvester equations once in every iteration. From the equivalence between the Krylov-subspace and Sylvester equation, it is obvious that both have comparable computational cost \cite{panzer2014model}. Due to the small size of $S_r$, $V_{r,\Omega}$ and $W_{r,\Omega}$ can be computed within admissible time, even for large-scale systems. Nevertheless, the computational cost of FLPORK and FLCURE is a fraction of that of FLIRKA if it does not converge in a few iterations. Unlike FLCURE, FLPORK constructs the ROM in a single-run, and thus its computational cost is slightly less than that of FLCURE if the same ROM is generated using both algorithms. If, however, the resultant ROM is not accurate enough, FLPORK needs to be restarted, and $V_{r,\Omega}$ or $W_{r,\Omega}$ needs to be recomputed all over again. This is where FLCURE stands out as it continues to reuse matrices from the previous interpolation data and accumulate these to construct the ROM. $V_{r,\Omega}$ or $W_{r,\Omega}$ in FLCURE is not required to be recomputed afresh, but it is only recursively updated. Thus the additional small-scale operations that FLCURE performs (unlike FLPORK) can potentially help to significantly bring down the computational cost in case the ROM is infeasible from an accuracy standpoint. FLBT, FLIRKA, FLPORK, and FLCURE all require the computation of matrix logarithm (\ref{e4}) that is also expensive in a large-scale setting. This needs to be approximated before solving the Lyapunov equations (\ref{e2}) and (\ref{e3}), or the Sylvester equation (\ref{666}). As mentioned before, the matrix logarithm (\ref{e4}) can be approximated by using the Krylov subspace-based method in \cite{benner2016frequency}, and $V_c^{(i)}$ in FLCURE can be used both for the computation of $V_{c,\Omega}^{(i)}$ and the approximation of $\mathscr{F}(A)$. Although not recognized in \cite{vuillemin2013h2} and \cite{zulfiqar2020frequency}, the relation proved in this paper, i.e., (\ref{e14}), can also be used by FLIRKA and FLPORK to perform both of these tasks, and some computational cost can be saved. Lastly, the stopping criteria of FLCURE only involve small-scale operations. However, if the interpolation data is selected using an optimization toolbox, the computational cost may be dominated by the optimization routine if it does not converge quickly. Therefore, we suggest using the heuristic selection procedure detailed earlier, and we will show in the next section that it produces promising results.
\section{Numerical Examples}
In this section, we test FLCURE on two numerical examples and compare its performance with FLBT and FLIRKA. We consider medium-scale systems so that the exact computation of $P_\Omega$ and $Q_\Omega$ remain possible for a fair comparison with FLBT. A desired tolerance of error in terms of $\mathcal{H}_{2,\Omega}$ is set in FLCURE, and it adaptively decides the order based on that. The interpolation data is selected automatically based on the heuristic selection criteria detailed in the last section, i.e., IRKA is initialized randomly, and the interpolation data is updated for $10$ iterations, which is then used to initialize FLIRA, and the interpolation data is again updated for $10$ iterations. The final interpolation data is used in FLCURE both for MOR and for computing the approximate Gramians. The size of ROM in each step of FLCURE is set to $2$. FLIRKA is also initialized via IRKA for a fair comparison, i.e., IRKA is initialized randomly, and the interpolation data is updated for $10$ iterations. The approximate Gramians generated by FLCURE, i.e., $\hat{P}_\Omega$ and $\hat{Q}_\Omega$, are used to perform approximate FLBT, which we refer to here as "CUREd-FLBT". The desired tolerance for both the MOR problem and the solutions of Lyapunov equations is set to $1\times 10^{-2}$, i.e., $||H(s)-H_c^{(i)}(s)||_{\mathcal{H}_{2,\Omega}}\leq1\times 10^{-2}$ and $\epsilon_{rel}^{(i)} \leq1\times 10^{-2}$.
\begin{figure}[b]
  \centering
  \includegraphics[width=6.5cm]{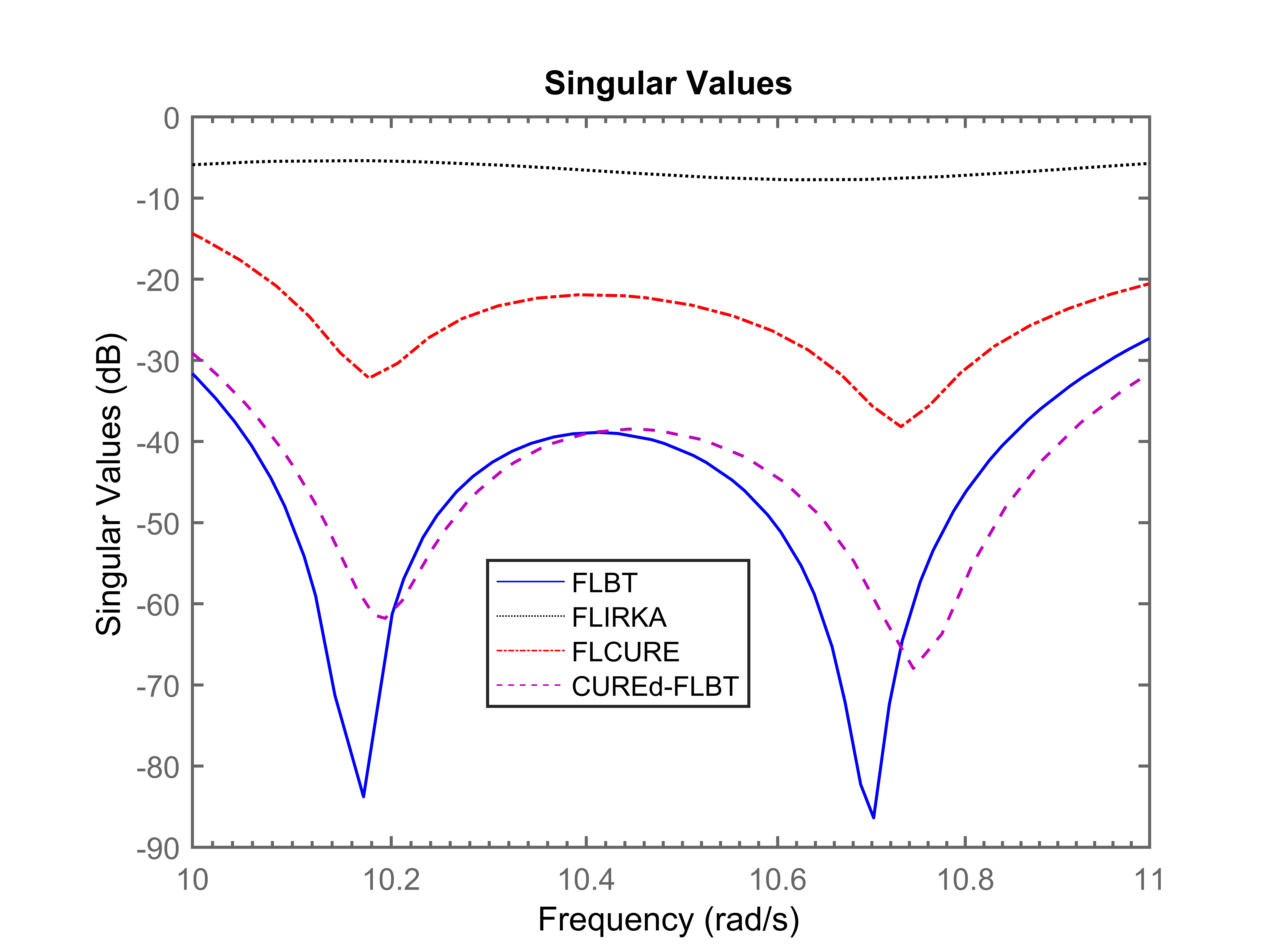}
  \caption{$\sigma\big(H(s)-H_r(s)\big)$ within $[10,11]$ rad/sec}\label{fig1}
\end{figure}

\textbf{Example 1: Clamped Beam}\\
This is a $348^{th}$-order SISO model taken from the benchmark collection of \cite{chahlaoui2005benchmark}. The frequency interval of interest is specified as $[10,11]$ rad/sec. FLCURE converged in only two iterations, both for MOR and the solutions of Lyapunov equations. The accuracy of the $4^{th}$-order ROMs generated by FLCURE and CUREd-FLBT is compared with that produced by FLBT and FLIRKA in Table \ref{tab2}. It can be seen that FLCURE ensures supreme accuracy. The sigma plot of $H(s)-H_r(s)$ within $[10,11]$ rad/sec, which is an effective visual depiction of the $\mathcal{H}_\infty$-norm error for the SISO systems, is plotted in Figure \ref{fig1}. As expected, FLBT yields the best results in $\mathcal{H}_\infty$-norm sense, and CUREd-FLBT compares well with FLBT.
\begin{table}[!t]
\centering
\caption{$||H(s)-H_r(s)||_{\mathcal{H}_{2,\Omega}}$}\label{tab2}
\begin{tabular}{|c|c|c|c|}
\hline
FLBT&FLIRKA&FLCURE&CUREd-FLBT\\ \hline
$0.0071$&$0.2675$&$5.0161\times 10^{-4}$&0.0059\\ \hline
\end{tabular}
\end{table}

\begin{figure}[t]
  \centering
  \includegraphics[width=6.5cm]{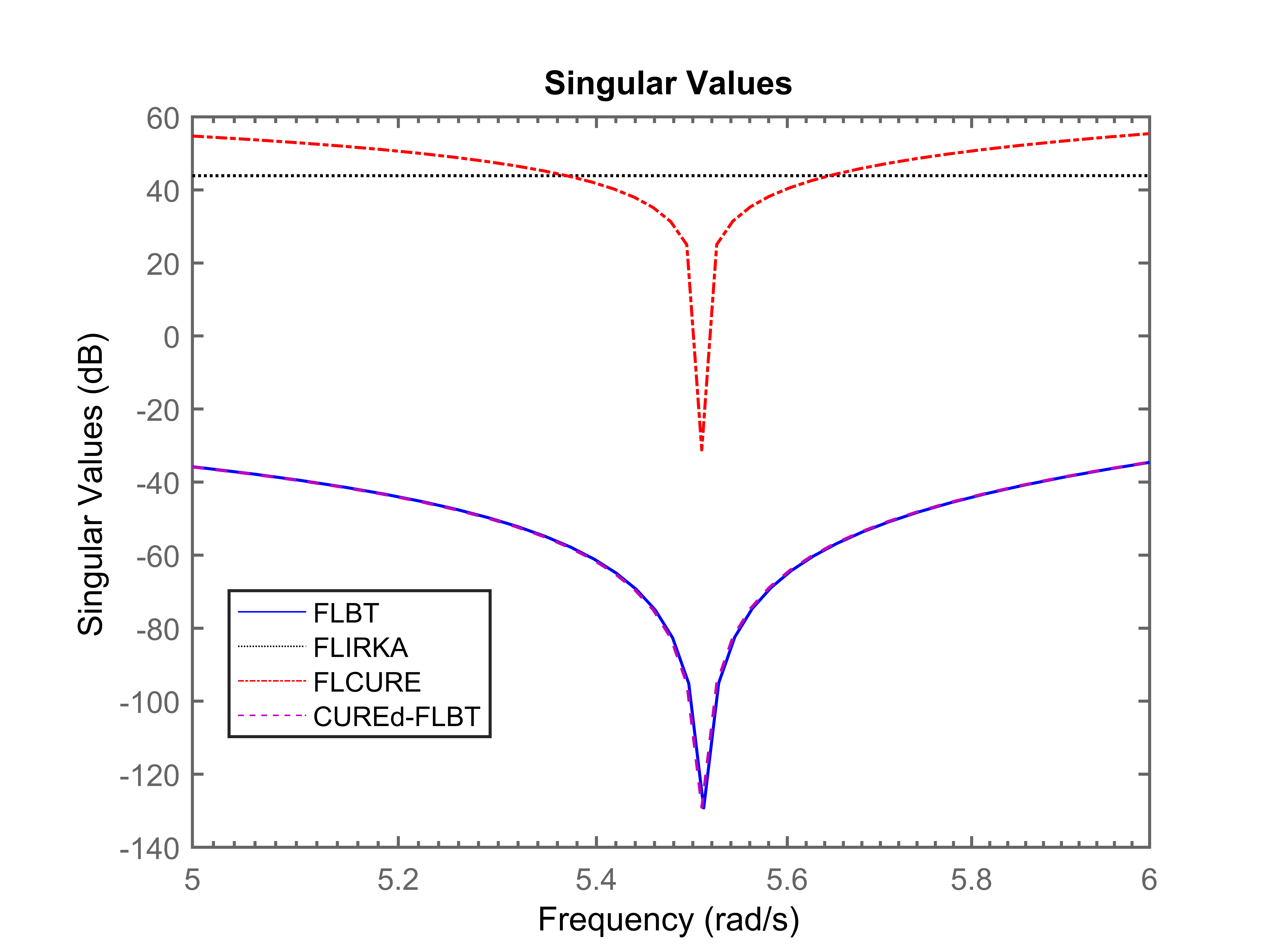}
  \caption{$\sigma\big(H(s)-H_r(s)\big)$ (Output 1) within $[5,6]$ rad/sec}\label{fig2}
\end{figure}

\textbf{Example 2: CD Player}\\
This is a $120^{th}$-order MIMO model taken from the benchmark collection of \cite{chahlaoui2005benchmark}. The frequency interval of interest is specified as $[5,6]$ rad/sec. Again, FLCURE converged in only two iterations, both for MOR and the solutions of Lyapunov equations. The accuracy of the $4^{th}$-order ROMs generated by FLCURE and CUREd-FLBT is compared with that produced by FLBT and FLIRKA in Table \ref{tab3}. It can be seen that FLCURE ensures supreme accuracy. The sigma plot of $H(s)-H_r(s)$ for the first output within $[5,6]$ rad/sec is plotted in Figure \ref{fig2}. As expected, FLBT yields the best results in $\mathcal{H}_\infty$-norm sense, and CUREd-FLBT compares well with FLBT.
\begin{table}[!t]
\centering
\caption{$||H(s)-H_r(s)||_{\mathcal{H}_{2,\Omega}}$}\label{tab3}
\begin{tabular}{|c|c|c|c|}
\hline
FLBT&FLIRKA&FLCURE&CUREd-FLBT\\ \hline
$0.0114$&$92.6295$&$0.0058$&0.0054\\ \hline
\end{tabular}
\end{table}
\section{Conclusion}
In this paper, we present a cumulative scheme for MOR in limited frequency interval that adaptively constructs the ROM such that the error decays monotonically after each iteration irrespective of the choice of interpolation points and tangential directions. Moreover, the algorithms also provide approximate frequency-limited Gramians of the system that can be used for computing approximate frequency-limited balanced realization. The numerical results validate the theory developed in the paper.
\section*{Acknowledgment}
This work is supported in part by National Natural Science Foundation of China under Grant (No. $61873336$, $61873335$), in part by the National Key Research and Development Program (No. $2020$YFB $1708200$), in part by the Foreign Expert Program (No. $20$WZ$2501100$) granted by the Shanghai Science and Technology Commission of  Shanghai Municipality (Shanghai Administration of Foreign Experts Affairs), in part by $111$ Project (No. D$18003$) granted by the State Administration of Foreign Experts Affairs, and  in part by the Fundamental Research Funds for the Central Universities under Grant (No. FRF-BD-19-002A).


\begin{thebibliography}{10}
\providecommand{\url}[1]{\texttt{#1}}
\providecommand{\urlprefix}{URL }
\expandafter\ifx\csname urlstyle\endcsname\relax
  \providecommand{\doi}[1]{doi:\discretionary{}{}{}#1}\else
  \providecommand{\doi}{doi:\discretionary{}{}{}\begingroup
  \urlstyle{rm}\Url}\fi

\bibitem{ahmad2010krylov}
M.~I. Ahmad, I.~Jaimoukha, and M.~Frangos, \textit{Krylov subspace restart
  scheme for solving large-scale Sylvester equations}, \textit{Proceedings of
  the 2010 American Control Conference}, IEEE, 5726--5731.

\bibitem{balakrishnan2001efficient}
V.~Balakrishnan, Q.~Su, and C.-K. Koh, \textit{Efficient balance-and-truncate
  model reduction for large scale systems}, \textit{Proceedings of the 2001
  American Control Conference.(Cat. No. 01CH37148)}, vol.~6, IEEE, 4746--4751.

\bibitem{beattie2009trust}
C.~A. Beattie and S.~Gugercin, \textit{A trust region method for optimal $\mathcal{H}_2$
  model reduction}, \textit{Proceedings of the 48h IEEE Conference on Decision
  and Control (CDC) held jointly with 2009 28th Chinese Control Conference},
  IEEE, 5370--5375.

\bibitem{beattie2014model}
C.~A. Beattie and S.~Gugercin, \textit{Model reduction by rational
  interpolation}, Model Reduction and Algorithms: Theory and Applications, P.
  Benner, A. Cohen, M. Ohlberger, and K. Willcox, eds., Comput. Sci. Engrg
  \textbf{15} (2014), 297--334.

\bibitem{benner2016frequency}
P.~Benner, P.~K{\"u}rschner, and J.~Saak, \textit{Frequency-limited balanced
  truncation with low-rank approximations}, SIAM Journal on Scientific
  Computing \textbf{38} (2016), no.~1, A471--A499.

\bibitem{benner2005dimension}
P.~Benner, V.~Mehrmann, and D.~C. Sorensen, \textit{Dimension reduction of
  large-scale systems}, vol.~45, Springer, 2005.

\bibitem{benner2017model}
P.~Benner et~al., \textit{Model reduction and approximation: theory and
  algorithms}, vol.~15, SIAM, 2017.

\bibitem{chahlaoui2005benchmark}
Y.~Chahlaoui and P.~Van~Dooren, \textit{Benchmark examples for model reduction
  of linear time-invariant dynamical systems}, \textit{Dimension Reduction of
  Large-Scale Systems}, Springer, 2005. 379--392.

\bibitem{chow2013power}
J.~H. Chow, \textit{Power system coherency and model reduction}, vol.~84,
  Springer, 2013.

\bibitem{gawronski1990model}
W.~Gawronski and J.-N. Juang, \textit{Model reduction in limited time and
  frequency intervals}, International Journal of Systems Science \textbf{21}
  (1990), no.~2, 349--376.

\bibitem{ghafoor2008survey}
A.~Ghafoor and V.~Sreeram, \textit{A survey/review of frequency-weighted
  balanced model reduction techniques}, Journal of Dynamic Systems,
  Measurement, and Control \textbf{130} (2008), no.~6.

\bibitem{gugercin2008h_2}
S.~Gugercin, A.~C. Antoulas, and C.~Beattie, \textit{$\mathcal{H}_2$ model
  reduction for large-scale linear dynamical systems}, SIAM journal on matrix
  analysis and applications \textbf{30} (2008), no.~2, 609--638.

\bibitem{haider2018frequency}
S.~Haider et~al., \textit{Frequency interval gramians based structure
  preserving model order reduction for second order systems}, Asian Journal of
  Control \textbf{20} (2018), no.~2, 790--801.

\bibitem{imran2015frequency}
M.~Imran and A.~Ghafoor, \textit{A frequency limited interval gramians-based
  model reduction technique with error bounds}, Circuits, Systems, and Signal
  Processing \textbf{34} (2015), no.~11, 3505--3519.

\bibitem{imran2015model}
M.~Imran and A.~Ghafoor, \textit{Model reduction of descriptor systems using
  frequency limited gramians}, Journal of the Franklin Institute \textbf{352}
  (2015), no.~1, 33--51.

\bibitem{imran2016model}
M.~Imran and A.~Ghafoor, \textit{Model reduction of generalized non-singular
  systems using limited frequency interval gramians}, IMA Journal of
  Mathematical Control and Information \textbf{33} (2016), no.~2, 333--347.

\bibitem{imran2017frequency}
M.~Imran and A.~Ghafoor, \textit{Frequency limited model reduction techniques
  with error bounds}, IEEE Transactions on Circuits and Systems II: Express
  Briefs \textbf{65} (2017), no.~1, 86--90.

\bibitem{jazlan2016frequency}
A.~Jazlan et~al., \textit{Frequency interval balanced truncation of
  discrete-time bilinear systems}, Cogent Engineering \textbf{3} (2016), no.~1,
  1203082.

\bibitem{jiang2018model}
Y.-L. Jiang, Z.-Z. Qi, and P.~Yang, \textit{Model order reduction of linear
  systems via the cross gramian and SVD}, IEEE Transactions on Circuits and
  Systems II: Express Briefs \textbf{66} (2018), no.~3, 422--426.

\bibitem{konkel2014posteriori}
Y.~Konkel et~al., \textit{A posteriori error bounds for Krylov-based fast
  frequency sweeps of finite-element systems}, IEEE transactions on magnetics
  \textbf{50} (2014), no.~2, 441--444.

\bibitem{kurschner2018balanced}
P.~K{\"u}rschner, \textit{Balanced truncation model order reduction in limited
  time intervals for large systems}, Advances in Computational Mathematics
  \textbf{44} (2018), no.~6, 1821--1844.

\bibitem{li2002low}
J.-R. Li and J.~White, \textit{Low rank solution of Lyapunov equations}, SIAM
  Journal on Matrix Analysis and Applications \textbf{24} (2002), no.~1,
  260--280.

\bibitem{moore1981principal}
B.~Moore, \textit{Principal component analysis in linear systems:
  Controllability, observability, and model reduction}, IEEE transactions on
  automatic control \textbf{26} (1981), no.~1, 17--32.

\bibitem{obinata2012model}
G.~Obinata and B.~D. Anderson, \textit{Model reduction for control system
  design}, Springer Science \& Business Media, 2012.

\bibitem{panzer2014model}
H.~K. Panzer, \textit{Model order reduction by Krylov subspace methods with
  global error bounds and automatic choice of parameters}, Ph.D. thesis,
  Technische Universit{\"a}t M{\"u}nchen, 2014.

\bibitem{penzl1999cyclic}
T.~Penzl, \textit{A cyclic low-rank smith method for large sparse Lyapunov
  equations}, SIAM Journal on Scientific Computing \textbf{21} (1999), no.~4,
  1401--1418.

\bibitem{petersson2013nonlinear}
D.~Petersson, \textit{A nonlinear optimization approach to $\mathcal{H}_2$-optimal modeling
  and control}, Ph.D. thesis, Link{\"o}ping University, 2013.

\bibitem{petersson2014model}
D.~Petersson and J.~L{\"o}fberg, \textit{Model reduction using a
  frequency-limited $\mathcal{H}_2$-cost}, Systems \& Control Letters \textbf{67} (2014),
  32--39.

\bibitem{rommes2006efficient}
J.~Rommes and N.~Martins, \textit{Efficient computation of multivariable
  transfer function dominant poles using subspace acceleration}, IEEE
  transactions on power systems \textbf{21} (2006), no.~4, 1471--1483.

\bibitem{rook1996modal}
T.~E. Rook and R.~Singh, \textit{Modal truncation issues in synthesis
  procedures for vibratory power flow and dissipation}, The Journal of the
  Acoustical Society of America \textbf{99} (1996), no.~4, 2158--2166.

\bibitem{schilders2008model}
W.~H. Schilders, H.~A. Van~der Vorst, and J.~Rommes, \textit{Model order
  reduction: theory, research aspects and applications}, vol.~13, Springer,
  2008.

\bibitem{shaker2013generalized}
H.~R. Shaker, \textit{Generalized frequency-interval balanced model reduction
  method}, \textit{52nd IEEE Conference on Decision and Control}, IEEE,
  5546--5551.

\bibitem{shaker2006frequency}
H.~R. Shaker, M.~Samavat, and A.~A. Ghareveisi, \textit{Frequency domain
  stochastic balanced truncation: An accuracy enhanced large scale model
  reduction technique}, \textit{2006 IEEE Conference on Computer Aided Control
  System Design, 2006 IEEE International Conference on Control Applications,
  2006 IEEE International Symposium on Intelligent Control}, IEEE, 3003--3006.

\bibitem{toor2018improved}
H.~I. Toor et~al., \textit{Improved frequency limited model reduction},
  \textit{2018 IEEE Conference on Systems, Process and Control (ICSPC)}, IEEE,
  17--22.

\bibitem{van2008h2}
P.~Van~Dooren, K.~A. Gallivan, and P.-A. Absil, \textit{$\mathcal{H}_2$-optimal model
  reduction of MIMO systems}, Applied Mathematics Letters \textbf{21} (2008),
  no.~12, 1267--1273.

\bibitem{vuillemin2014frequency}
P.~Vuillemin, \textit{Frequency-limited model approximation of large-scale
  dynamical models}, Ph.D. thesis, Universit{\'e}t de Toulouse, 2014.

\bibitem{vuillemin2013h2}
P.~Vuillemin, C.~Poussot-Vassal, and D.~Alazard, \textit{$\mathcal{H}_2$ optimal and
  frequency limited approximation methods for large-scale LTI dynamical
  systems}, IFAC Proceedings Volumes \textbf{46} (2013), no.~2, 719--724.

\bibitem{wilson1970optimum}
D.~Wilson, \textit{Optimum solution of model-reduction problem},
  \textit{Proceedings of the Institution of Electrical Engineers}, vol. 117,
  IET, 1161--1165.

\bibitem{wolf2014h}
T.~Wolf, \textit{$\mathcal{H}_2$ pseudo-optimal model order reduction}, Ph.D.
  thesis, Technische Universit{\"a}t M{\"u}nchen, 2014.

\bibitem{wolf2016adi}
T.~Wolf, H.~K. Panzer, and B.~Lohmann, \textit{ADI iteration for Lyapunov
  equations: A tangential approach and adaptive shift selection}, Applied
  Numerical Mathematics \textbf{109} (2016), 85--95.

\bibitem{zulfiqar2019finite}
U.~Zulfiqar, V.~Sreeram, and X.~Du, \textit{Finite-frequency power system
  reduction}, International Journal of Electrical Power \& Energy Systems
  \textbf{113} (2019), 35--44.

\bibitem{zulfiqar2020frequency}
U.~Zulfiqar, V.~Sreeram, and X.~Du, \textit{Frequency-limited pseudo-optimal
  rational Krylov algorithm for power system reduction}, International Journal
  of Electrical Power \& Energy Systems \textbf{118} (2020), 105798.

\bibitem{zulfiqar2016new}
U.~Zulfiqar et~al., \textit{A new frequency-limited interval gramians-based
  model order reduction technique}, IEEE Transactions on Circuits and Systems
  II: Express Briefs \textbf{64} (2016), no.~6, 680--684.

\bibitem{zulfiqar2020time}
U.~Zulfiqar et~al., \textit{Time/frequency-limited positive-real truncated
  balanced realizations}, IMA Journal of Mathematical Control and Information
  \textbf{37} (2020), no.~1, 64--81.

\end{thebibliography}

\end{document}